\documentclass[preprint]{aastex631}

\usepackage{
graphicx,
color,
natbib,
}

\newcommand{\jy}{JY$_{\rm 31}$}
\newcommand{\os}{OS$_{\rm 393}$}
\newcommand{\hz}{HZ$_{\rm 102}$}
\newcommand{\hk}{HK$_{\rm 103}$}
\newcommand{\pn}{PN$_{\rm 70}$}

\newcommand{\nh}{\emph{New Horizons}\ }
\newcommand{\hst}{\emph{HST}\ }
\newcommand{\onebyone}{1$\times$1\ }
\newcommand{\fourbyfour}{4$\times$4\ }

\received{August 15, 2021}
\revised{January 2, 2022}
\accepted{January 15, 2022}
\submitjournal{Planetary Science Journal}

\begin{document}

\title{High Resolution Search for KBO Binaries from \nh}

\correspondingauthor{H. A. Weaver}
\email{hal.weaver@jhuapl.edu}

\author[0000-0003-0951-7762]{H. A. Weaver}
\affiliation{Johns Hopkins University Applied Physics Laboratory,
11100 Johns Hopkins Road,
Laurel, MD 20723-6099, USA}

\author[0000-0003-0333-6055]{S. B. Porter}
\affiliation{Southwest Research Institute,
1050 Walnut Street, Suite 300,
Boulder, CO 80302, USA}

\author{J. R. Spencer}
\affiliation{Southwest Research Institute,
1050 Walnut Street, Suite 300,
Boulder, CO 80302, USA}


\author{The \nh Science Team}

\begin{abstract}

Using the \nh LORRI camera, we searched for satellites
near five Kuiper belt objects (KBOs): four cold classicals (CCs: 2011~\jy, 2014~\os,
2014~\pn, 2011~\hz) and one scattered disk object (SD: 2011~ \hk).
These objects were observed 
at distances of $0.092$-$0.290$~au from the \nh spacecraft, 
achieving spatial resolutions 
of \mbox{$136$-$430$ km} (resolution is $\sim$2 camera pixels), much higher than possible from any other facilities.
Here we report that CC 2011~\jy\ is a binary system with roughly equal brightness components, 
CC 2014~\os\ is likely an equal brightness binary system, while the three other KBOs
did not show any evidence of binarity.
The \jy\ binary has a semi-major axis of \mbox{$198.6 \pm 2.9$ km}, 
an orbital inclination of \mbox{$61\fdg 34 \pm 1\fdg 34$},
and an orbital period of \mbox{$1.940 \pm 0.002$ d.}
The \os\ binary objects have an apparent separation of $\sim$150~km,
making \jy\ and \os\ the tightest KBO binary systems ever resolved.
Both \hk\ and \hz\ were detected with SNR$\approx$10, and our observations
rule out equal brightness binaries with separations larger than $\sim$430~km and $\sim$260~km, respectively.
The spatial resolution for \pn\ was \mbox{$\sim$200 km}, but this object had
\mbox{SNR$\approx$2.5--3,} which limited our ability to probe its binarity.
The binary frequency for the CC binaries probed
in our small survey (67\%, not including \pn) is consistent with the high binary frequency suggested 
by larger surveys of CCs \citep{fraser:2017, fraser:2017correction, noll:tno2020}
and recent planetesimal formation models \citep{Nesvorny:psj2021},
but we extend the results to smaller orbit semi-major axes and
smaller objects than previously possible.

\end{abstract}



\section{Introduction} \label{sec:intro}
The \nh ({\it NH}) spacecraft has been traversing the Kuiper belt since 2015.
The close flyby of Pluto on 2015~July~14 at a distance of $\sim$12,500~km revolutionized our understanding of that
dwarf planet and its satellites \citep[cf.,][]{stern:2018araa}.
On 2019~January~1, the \nh spacecraft passed within 3600~km of the cold classical (CC) 
Kuiper belt object (KBO) (486958) Arrokoth, conducting 
ultraviolet, visible, near-infrared, and radio investigations with unprecedented resolution and sensitivity
that have revealed unique results on planetesimal formation and evolution in the distant solar system 
\citep{stern:2019arrokoth, grundy:2020arrokoth, mckinnon:2020arrokoth, spencer:2020arrokoth}.
More generally, \nh has enabled observations of dozens of KBOs at unique geometries, 
including at large solar phase angles not possible from the inner solar system 
\citep[e.g.,][]{porter:2016jr1, verbiscer:2019dkbos, verbiscer:submitted} 
and at distances from the spacecraft that provide higher spatial resolution than available from Earth, or Earth-orbiting, facilities.
Here we describe how this latter capability has been exploited to provide sensitive searches for possible satellites
of KBOs passing within 0.3~au of the \nh spacecraft when the spatial resolution exceeded that
available from the {\it Hubble Space Telescope} ({\it HST}) by a factor $\geq$6.
The \nh flyby observations of Arrokoth revealed it to be a contact binary with two distinct lobes,
but sensitive searches did not uncover any other satellites larger than a few hundred meters in diameter
within a few hundred kilometers of the main body \citep{stern:2019arrokoth}.

These \nh high resolution KBO satellite searches were conducted using 
the LOng Range Reconnaissance Imager (LORRI),
which is a panchromatic \mbox{(360--910 nm),} narrow angle \mbox{(field of view$=$FOV$=$0.291\arcdeg),} 
high spatial resolution \mbox{(pixel scale$=$IFOV$=$1\farcs02),}
Ritchey-Chr\'{e}tien telescope with a CCD in its focal plane that is used 
for both science observations and optical navigation \citep{cheng:2008ssr, weaver:2020pasp}.
For the highest resolution LORRI observations, like those discussed here, all optically active pixels are read out from the CCD 
(``1$\times$1'' format with 1024$\times$1024 optically active pixels).
When used in \onebyone format, and employing a typical exposure time of 150~ms to prevent pointing smear, LORRI can achieve 
a signal-to-noise ratio (SNR) of $\sim$5 for unresolved targets with \mbox{$V \leq 12.6$.}
However, the pixels can be also be re-binned by a factor of 4 in each direction (i.e., column and row directions)
during CCD readout (``4$\times$4'' format with 256$\times$256 optically active pixels), which reduces the data volume by a factor of 16 and 
results in an effective pixel size of IFOV$=$4\farcs08.
By employing \mbox{$4 \times 4$} mode, a special
spacecraft tracking mode, exposure times $\geq$30~s, and co-addition of $\sim$100 images, 
LORRI can detect unresolved targets down to \mbox{$V \approx 22$} with \mbox{SNR $\approx$ 5.}
This latter capability has been used extensively to measure KBO phase curves and light curves over wide 
ranges of solar phase angles \citep{porter:2016jr1, verbiscer:2019dkbos, verbiscer:submitted}, 
which can only be achieved by a facility in the outer solar system.

The five KBOs discussed here were observed in both \onebyone and \fourbyfour format.
The \onebyone observations were conducted primarily to search for potential satellites at high spatial resolution
and are the main subject of this paper.
Preliminary results from this program were discussed in \citet{weaver:agu2019},
and the discovery that \jy\ is the tightest known KBO binary system was announced in \citet{porter:dps2020}.
The \fourbyfour observations were conducted to measure light curves, phase curves, and to refine the
orbit solutions for these KBOs via astrometry; the details of the \fourbyfour observations are discussed
in a separate paper \citep{porter:submitted}.

For the LORRI \onebyone observations, hundreds of images had to be co-added, 
and longer than usual exposure times (500~ms) were needed 
to reach the sensitivity required to detect these small and faint KBOs,
which had \mbox{$V \approx 14.5-17.3$.}
We employed a special windowing technique to restrict
the sizes of the images sent back to the Earth, which reduced the 
downlinked data volumes to $\sim$20\% of their full-size values.
The special techniques applied to these high resolution
KBO observations are described further below.

\newpage

\section{Observations} \label{sec:obs}
We observed five KBOs, which included four CCs and one scattered disk object (SD).
All of these KBOs were discovered during dedicated searches 
for potential \nh observational targets:
three (2011~\hz, 2011~\jy, 2011~\hk) were discovered during a ground-based survey program
and two (2014~\os, 2014~\pn) were discovered during a search with {\it HST} \citep{spencer:2015}.
Arrokoth, the target of a \nh mission close flyby campaign,
was also discovered by that same \hst program.
\os\ and \pn\ were potential close flyby \nh targets,
but they required much more fuel to reach than Arrokoth and had later arrival times, so they
were dropped from consideration.

Some key properties of these KBOs are provided in Table~\ref{tab:properties}.
The dynamical class assignments follow the standard conventions \citep{gladman:tno2008,noll:icarus2008}.
The orbital parameters and absolute visual magnitudes ($H_{0}$) are taken from
the Minor Planet Center.
The geometric albedos have not been measured for these KBOs, so
we used representative values for their dynamical classes.
For \hk\ we use the average geometric albedo derived from
{\it Herschel} observatory observations of SD KBOs \citep{lacerda:2014}.
For the CC KBOs, we use the average geometric albedo derived
from {\it Herschel} observatory observations of CCs \citep{vilenius:2014}.
We note that the reported geometric albedo of CC Arrokoth is
$0.21_{-0.04}\!\!\!\!\!\!\!\!\!\!\!^{+0.05}$ \citep{hofgartner:2021}, which is
slightly larger than, but consistent within the error bars, the average CC value.
The approximate effective diameters (d$_{N}$) are derived from the listed
$H_{0}$ values and geometric albedos.
Further details on these KBOs are discussed elsewhere 
\citep{verbiscer:2019dkbos,porter:submitted}.
All of the KBOs discussed here are small ($\la$100~km in diameter) 
and require large telescopes ($\ga$6.5~m) when observed from the vicinity of the Earth.
 
\begin{deluxetable*}{lccccccr}
\tablecaption{KBO Properties \label{tab:properties}}
\tablewidth{0pt}
\tablehead{
\colhead{Target} & \colhead{Class} & \colhead{a} & \colhead{e} & \colhead{i} 
& \colhead{H$_{0}$} & \colhead{p$_{V}$} & \colhead{d$_{N}$} \\
& & \colhead{(au)} & & \colhead{(deg)} & & & \colhead{(km)}
}
\startdata
2011 \hk & SD & 53.07 & 0.311 & 6.43 
& 8.4 & 0.08 & 99 \\
2011 \jy & CC & 43.95 & 0.059 & 2.61 
& 8.8 & 0.15 & 60 \\
2011 \hz & CC & 43.18 & 0.0061 & 2.39 
& 9.2 & 0.15 & 50 \\
2014 \os & CC & 43.94 & 0.032 & 3.81 
& 10.1 & 0.15 & 33 \\
2014 \pn & CC& 44.04 & 0.050 & 4.12 
& 10.3 & 0.15 & 30 \\
\enddata
\tablecomments{``Target'' is the IAU KBO designation.
``Class'' refers to the KBO dynamical class:
``CC'' is short for ``cold classical'' and ``SD'' is short for ``scattered disk''.
``a'', ``e'', ``i'' are the semi-major axis, eccentricity, and inclination angle,
respectively, of the KBO's orbit.
``H$_0$'' is the KBO's absolute $V$-mag from the Minor Planet Center.
``p$_{V}$'' is the KBO's assumed $V$-band geometric albedo 
based on its dynamical class.
``d$_{N}$'' is the KBO's estimated diameter.
See the text for further details.
}
\end{deluxetable*}

The LORRI \onebyone observations of these KBOs posed a number of technical challenges,
principally involving sensitivity limitations and downlink data volume (DLDV) considerations.
Pointing drift during observations limits the sensitivity achieved by individual images; exposures
that are too long allow the target to move across multiple pixels thereby lowering the signal
accumulated in the peak pixel and reducing the SNR.
Pointing control on the \nh spacecraft is established by firing hydrazine thrusters and monitoring
the response autonomously using star locations derived from a \nh attitude star tracker 
(AST)\footnote{There are two ASTs for redundancy, but only one is active at a time.
Both ASTs employ CCD detectors.} 
and rates of motion derived from an inertial measurement unit 
(IMU)\footnote{There are two IMUs for redundancy, but only one is active at a time.
Each IMU is comprised of three laser ring gyros mounted orthogonally to each other.}.
During these \onebyone KBO observations, the pointing was controlled to lie within
a ``deadband'' of $\pm$250~$\mu$rad ($\pm$52\arcsec) centered on the commanded 
target location in the LORRI field-of-view (FOV), always the center of the FOV 
(i.e., the LORRI ``boresight'') for these cases.
The total rate of motion (summed over all 3 directions, including spacecraft roll) was controlled to be 
\mbox{$\leq$38 $\mu$rad s$^{-1}$} \mbox{($\leq$7\farcs8 s$^{-1}$).}
Whenever the pointing starts to breach these location or rate limits during an observation, the 
spacecraft's attitude control system (ACS) fires the thrusters to maintain the
pointing within these specified deadbands, typically at the rate of $\sim$6 thruster firings per minute.
Thus, the pointing drifts during an observation within a box whose 
width is $\pm$52\arcsec\ about the commanded location.
This drift is taken into account when shifting and co-adding multiple images to produce
more sensitive composites, as described later.
The \nh pointing control system \citep[see][]{rogers:2016} has been stable with essentially identical performance
over the entire mission duration.

\begin{figure}[h!]
\includegraphics[keepaspectratio,width=\linewidth]{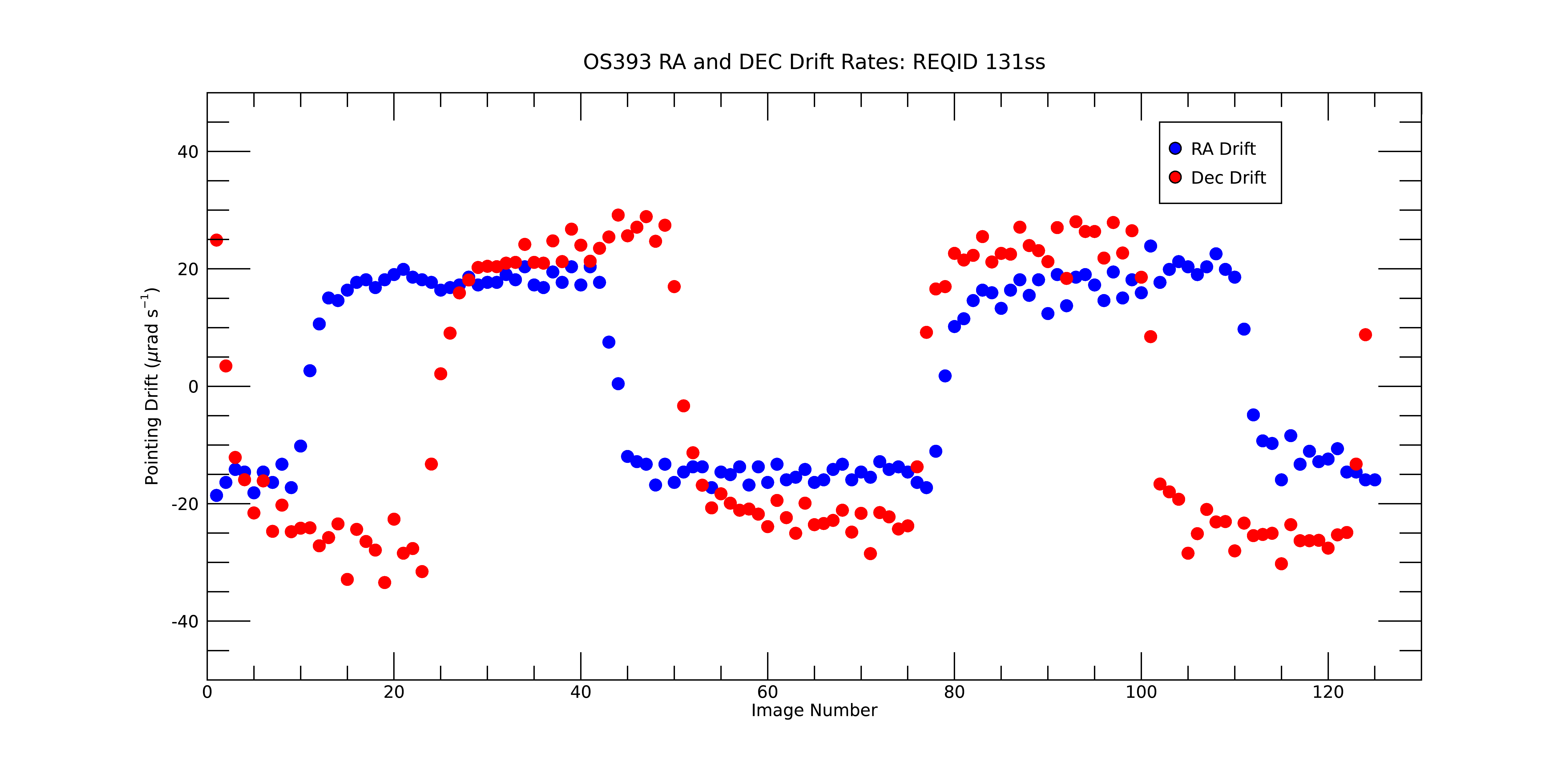}
\caption{RA and DEC drift rates are displayed for the 125 images taken during the
observations of \os\ for REQID 131ss.
The pointing drifts within a deadband of $\pm$250~$\mu$rad ($\pm$52\arcsec)
centered on the commanded target ephemeris location.
Hydrazine thrusters fire to maintain the pointing within the deadband;
the times of thruster firings are associated with abrupt changes in either
the RA or DEC drift rates in this plot.
The mean total plane-of-sky drift rate for this set of observations was
\mbox{$27.34 \pm 4.51 \mu$rad s$^{-1}$} \mbox{($5.64 \pm 0.93$ arcsec s$^{-1}$),} 
which is typical for \nh observations.
}
\label{fig:drift}
\end{figure}

One important consideration for our program was the choice of an exposure time that maximizes sensitivity
while mitigating image smear caused by pointing drift. 
Typically, the pointing for observations using the control parameters
discussed above drifts at an average rate of 
\mbox{$\sim$25 $\mu$rad s$^{-1}$} \mbox{($\sim$5\arcsec\ s$^{-1}$).}
Given that the intrinsic point spread function (PSF) of LORRI is $\sim$2.5~pixels
(FWHM for the best-fit 2-dimensional gaussian; see \citet{weaver:2020pasp}), 
we usually pick exposure times of $\leq$150~ms to keep the image smear $\leq$1~pixel.
For these faint KBO targets, however, we increased the exposure time
to 500~ms, which improves the single image sensitivity by a factor $\sim$3 while
only modestly blurring the PSF.
The pointing drifts for the KBO observations discussed here are consistent with
what has been seen throughout the mission;
an example is provided in Figure~\ref{fig:drift}.

\begin{figure}[h!]
\includegraphics[keepaspectratio,width=0.9\linewidth]{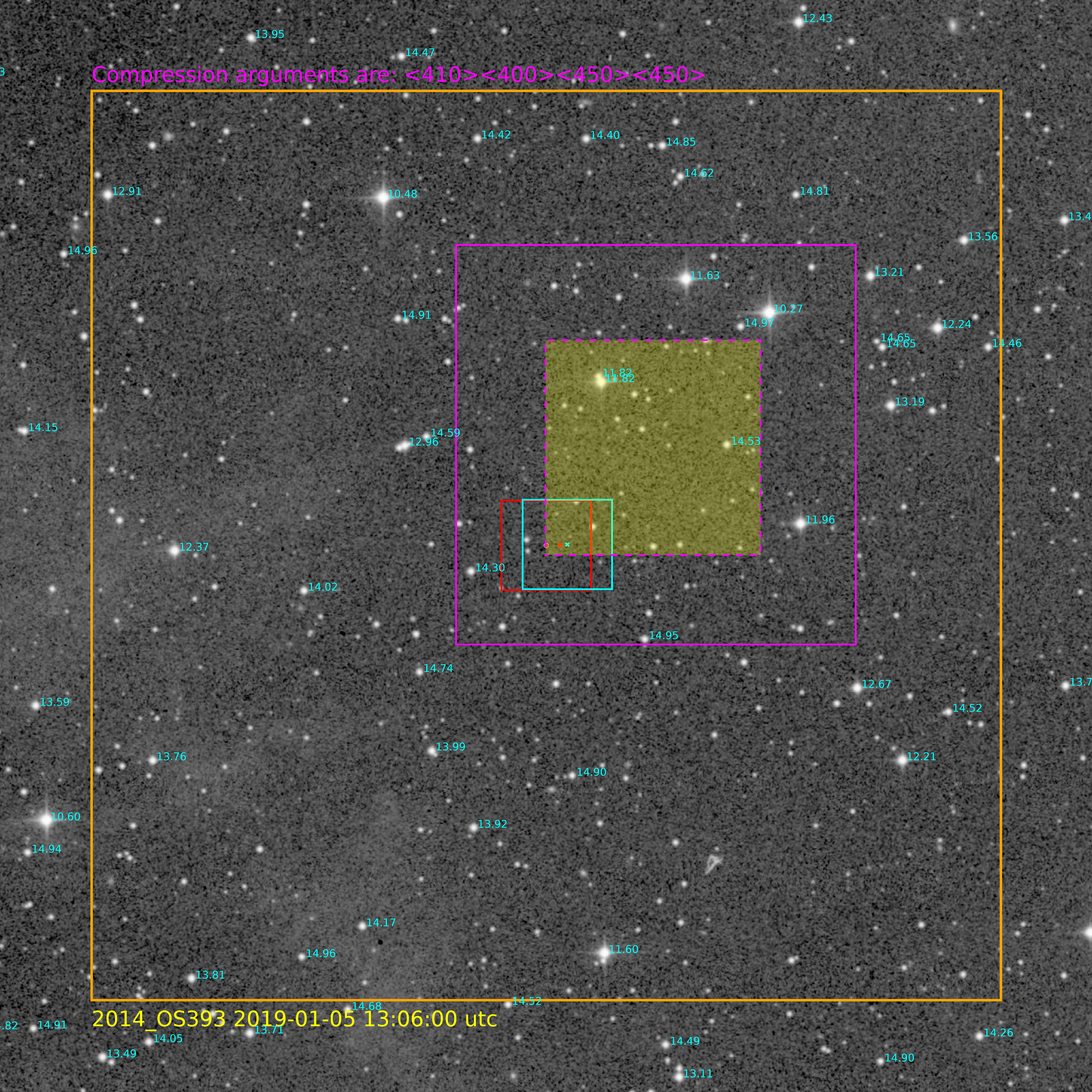}
\caption{Window selection details for one set (REQID$=$131ss) of observations of \os.
The background image is taken from the Palomar Digital Sky Survey (red filter, second generation reduction);
celestial north points straight down and celestial east points to the right.
The orange box is the LORRI FOV with X$_{\rm LORRI}$ (the direction of increasing CCD columns) pointing to the right
and Y$_{\rm LORRI}$ (the direction of increasing CCD rows) pointing up.
The magenta box is the requested window 
(i.e., the window specified by the command arguments listed in magenta just above the orange box). 
The dashed magenta box (with light yellow background highlighting) is the magenta box 
reduced by the size of the pointing deadband \mbox{(500~$\mu$rad $=$ 104\arcsec)} in each dimension.
We generally sized the image windows to capture both the KBO and at least one star with \mbox{$V \leq 13$}
within the dashed magenta box.
The red dot gives the LORRI boresight location, which is where the 
pointing control system commanded \os\ to be placed. 
The red box shows the size of the pointing deadband centered on the LORRI boresight. 
The red X shows the predicted \os\ location using the ephemeris used in planning the observations,
which is offset from the boresight by the absolute pointing error derived using astrometry from \fourbyfour
images taken near the time of the \onebyone observations.
%
The cyan X shows the predicted \os\ location using the new ephemeris generated based on the location of \os\
in the \fourbyfour images.
The cyan box is the pointing deadband centered on that location.
The $V$ magnitudes from APASS for stars in the UCAC4 catalog are indicated in cyan.
}
\label{fig:os-window}
\end{figure}

The individual images taken with an exposure time of 500~ms achieve a sensitivity limit (SNR$=$5) 
on unresolved targets of \mbox{$V \approx 13.6$}, which is approximately 1~mag fainter than
can be achieved using an exposure time of 150~ms.
However, the KBOs targeted in this program were predicted to have \mbox{$V =$ 15.7-16.1},
which requires co-adding at least 100 images to achieve detections of the fainter objects,
much less provide a sensitive search for satellites.
But downlinking hundreds of full LORRI \onebyone images would be prohibitively expensive;
a single compressed LORRI image produces a DLDV of $\sim$5~Mbits 
and requires $\sim$1~h of 70~m antenna time from NASA's Deep Space Network (DSN).
Thus, each image was windowed to reduce the DLDV by a factor of $\sim$5, 
while still retaining a large enough portion of the LORRI FOV to enable accurate 
astrometric solutions using stars captured in the windowed portion of image.
Since the KBOs themselves couldn't be detected in single images, producing an astrometric
solution based on the detected stars is critical for enabling accurate co-addition of the images.
(The pointing information from the spacecraft's ACS isn't accurate enough for this purpose.)
In practice, we chose observation times when at least one star with \mbox{$V \leq 13$} was within the
same window containing the targeted KBO, something that was achieved for the vast majority of the images collected
in this program. The only exception was \pn, which landed slightly outside the specified window
for 229 of the 500 images taken owing to the large spatial separation between the
KBO and the single bright nearby star.
The window selection details for a more typical set of observations
(in this case REQID 131ss for \os; ``REQID'' is one of the FITS keywords 
for LORRI images used as a target identifier) are provided in Figure~\ref{fig:os-window}.

For each KBO discussed here, \nh took four sets of 125 LORRI \onebyone images, 
with an exposure time of 500~ms for each image taken at 1 second cadence.
Thus, it took 125 seconds to complete each set. 
Two sets of 125 images, separated by $\sim$5-13 minutes, were taken the same day, 
then two more identical sets of 125 images each were taken 1 to 4 days later, 
depending on the target (see Table~\ref{tab:obs}). 
For all observations, \nh was tracking the KBO using the target's ephemeris, 
but the pointing drifted from image to image, as discussed above.
Owing to onboard data management considerations, which were exacerbated by the Arrokoth flyby
on New Year's Day 2019, the full KBO images had to be deleted from the solid state recorder 
(SSR)\footnote{There are two SSRs for redundancy, and the raw KBO images
had to be stored on SSR2, which is not normally powered.} before
the windowed portions could be downlinked.
The downlink of the windowed images from the LORRI \onebyone program was
not completed until April 2020.

\clearpage
\begin{longrotatetable}
\begin{deluxetable*}{cccccccccr}
\tabletypesize{\scriptsize}
\tablecaption{\nh KBO High Resolution Observations \label{tab:obs}}
\tablewidth{700pt}
\tablehead{
\colhead{Target} & \colhead{Dynamical} & \colhead{REQID} 
& \colhead{MET} & \colhead{Obs Date} 
& \colhead{$r$} & \colhead{$\Delta$} & \colhead{Phase}
& \colhead{SEA} & \colhead{Res}\\
& \colhead{Class} &
& \colhead{Range} & \colhead{(UTC)} 
& \colhead{(au) } &\colhead{(au)} & \colhead{(deg)}
& \colhead{(deg)} & \colhead{(km)}
}
\startdata
2011 \hk & SD & 123u 
& 396814919$-$396815043 & 2018-08-17 12:30:01$-$12:32:05
& 42.36 & 0.290 & 50.6 
& 129.1 & 215 \\
& & 123v 
& 396815819$-$396815943 & 2018-08-17 12:45:02$-$12:47:05
& & & 
& & \\
& & 123w 
& 397013099$-$397013223 & 2018-08-19 19:33:01$-$19:35:05
& 42.35 & 0.280 & 54.1
& 125.6 & 208\\
& & 123x
& 396815819$-$396815943 & 2018-08-19 19:48:01$-$19:50:05
& & & 
& & \\
2011 \jy & CC & 126oo 
& 398776019$-$398776143 & 2018-09-09 05:15:01$-$05:17:05
& 42.44 & 0.154 & 57.9 
& 121.9 & 114 \\
& & 126pp 
& 398776919$-$398777043 & 2018-09-09 05:30:01$-$05:32:05
& & & 
& & \\
& & 126qq 
& 398986019$-$398986143 & 2018-09-11 15:35:01$-$15:37:05
& 42.44 & 0.145 & 64.8
& 115.0 & 108\\
& & 126rr
& 398986919$-$398987043 & 2018-09-11 15:50:01$-$15:52:05
& & & 
& & \\
2011 \hz & CC & 129qq 
& 405876104$-$405876228 & 2018-11-30 09:29:46$-$09:31:50
& 43.15 & 0.185 & 47.1 
& 132.7 & 137 \\
& & 129rr 
& 405876494$-$405876618 & 2018-11-30 09:36:16$-$09:38:20
& & & 
& & \\
& & 129ss 
& 406219919$-$406220043 & 2018-12-04 09:00:01$-$09:02:05
& 43.15 & 0.168 & 56.0
& 123.8 & 125\\
& & 129tt
& 406220309$-$406220433 & 2018-12-04 09:06:31$-$09:08:35
& & & 
& & \\
2014 \os & CC & 131ss 
& 408999419$-$408999543 & 2019-01-05 13:05:01$-$13:07:05
& 43.33 & 0.0918 & 81.9 
& 98.0 & 68.2 \\
& & 131tt 
& 408999824$-$408999948 & 2019-01-05 13:11:46$-$13:13:50
& & & 
& & \\
& & 131qq 
& 409067559$-$409067683 & 2019-01-06 08:00:41$-$08:02:45
& 43.33 & 0.0922 & 85.9
& 94.0 & 68.5\\
& & 131rr
& 409067954$-$409068078 & 2019-01-06 08:07:16$-$08:09:20
& & & 
& & \\
2014 \pn & CC & 132v 
& 414396119$-$414396243 & 2019-03-09 00:10:01$-$00:12:05
& 43.89 & 0.139 & 59.6 
& 120.2 & 103 \\
& & 132w 
& 414396719$-$414396843 & 2019-03-09 00:20:01$-$00:22:05
& & & 
& & \\
& & 132x
& 414568919$-$414569043 & 2019-03-11 00:10:01$-$00:12:05
& 43.89 & 0.129 & 65.4
& 114.5 & 95.8\\
& & 132y
& 414569519$-$414569643 & 2019-03-11 00:20:01$-$00:22:05
& & & 
& & \\
\enddata
\tablecomments{``Target'' is the IAU KBO designation.
``Dynamical Class'' refers to the KBO dynamical class:
``CC'' is short for ``cold classical'' and ``SD'' is short for ``scattered disk''.
``REQID'' is one of the FITS keywords for LORRI images used as a target identifier;
for the actual FITS keywords, the REQIDs listed here are preceded with 
KALR\_HK103\_, KALR\_JY31\_, KALR\_HZ102\_, KDLR\_OS393\_, and K2LR\_PN70\_
for the KBOs in this program.
``MET'' stands for Mission Elapsed Time and is used for spacecraft timekeeping.
``Obs Date'' values are universal coordinated time (UTC) values for the mid-times of
the first and last images in the relevant sequence in the spacecraft frame.
``r'' is the target's heliocentric distance.
``$\Delta$'' is the target's distance from the spacecraft.
``Phase'' is to the sun-target-spacecraft angle (solar phase angle).
``SEA'' is the sun-spacecraft-target angle (solar elongation angle).
``Res'' is the projected distance in kilometers subtended by 1 LORRI pixel at the KBO.
The actual spatial resolution at the KBO is approximately two LORRI pixels.
}
\end{deluxetable*}
\end{longrotatetable}
\clearpage

\section{KBO Photometry} \label{sec:photometry}
We performed aperture photometry on all the LORRI \onebyone images (Table~\ref{tab:photometry}).
In all cases, we used a 5-pixel radius circular aperture to measure the signal from the KBO,
and we subtracted a background level derived from the mode of the signal contained within an annulus
with inner radius of 10 pixels and outer radius of 20 pixels.
The error in the net signal was calculated in the standard way, accounting for both the
photon statistics of the KBO signal and the variation of the signal in the background region.
The measured signals include contributions from both components of the binaries discussed
in the next section.

We converted LORRI signal rates to standard $V$ magnitudes
in the Johnson photometric system using:
\begin{equation}
\mathrm{
V = -2.5 \log (S/t_{exp}) +ZPT + COLOR - AC
}
\label{eqn:vmag}
\end{equation}
where $V$ is the magnitude in the standard Johnson $V$ band (i.e., specifies the target's flux at 5500~\AA),
S is the measured signal in the selected photometric aperture (DN), 
t$_{\mathrm{exp}}$ is the exposure time (0.5~s for all the images in this program),
$\mathrm{ZPT}$ is the photometric zero point (18.78 for \onebyone LLORRI images),
$\mathrm{COLOR}$ is a color correction term of 0.067 assuming an Arrokoth-like color \citep{grundy:2020arrokoth}
for all the KBOs, 
and $\mathrm{AC}$ is an aperture correction term to convert from the flux collected in a specified synthetic aperture 
to the total flux integrated over the LORRI PSF (0.1 was used here).

The two independent measurements on the same day are generally consistent (i.e., equal to each
other within their 1$\sigma$ errors).
However, several of the objects show significant changes in brightness from one observation date
to the other, indicating substantial lightcurve variations over that time.
More extensive lightcurve measurements for two of these KBOs (\hk\ and \jy) were obtained
as part of the LORRI \fourbyfour program \citep{verbiscer:2019dkbos}, and the $V$-mag values
reported here are consistent with those reported for the \fourbyfour program.

\begin{deluxetable*}{lrrr}
\tablecaption{\nh KBO Photometry \label{tab:photometry}}
\tablewidth{0pt}
\tablehead{
\colhead{Target} & \colhead{REQID} & \colhead{Signal (DN)} & \colhead{$V$}
}
\startdata
2011 \hk & 123u & $9.94 \pm 1.01$ & $15.59 \pm 0.11$ \\
& 123v & $8.95 \pm 0.96$ & $15.71 \pm 0.12$ \\
& 123uv & $9.40 \pm 0.90$ & $15.65 \pm 0.10$ \\
& 123w & $8.29 \pm 1.18$ & $15.79 \pm 0.15$ \\
& 123x & $8.36 \pm 0.96$ & $15.78 \pm 0.12$ \\
& 123wx & $8.39 \pm 0.84$ & $15.78 \pm 0.11$ \\
2011 \jy & 126oo & $27.62 \pm 1.486$ & $14.48 \pm 0.058$ \\
& 126pp & $25.00 \pm 1.278$ & $14.59 \pm 0.055$ \\
& 126oopp & $25.75 \pm 1.231$ & $14.56 \pm 0.052$ \\
& 126qq & $21.53 \pm 1.432$ & $14.75 \pm 0.072$ \\
& 126rr & $23.59 \pm 1.423$ & $14.65 \pm 0.065$ \\
& 126qqrr & $21.45  \pm 1.240$ & $14.76 \pm 0.063$ \\
2011 \hz & 129qq & $9.52 \pm 1.15$ & $15.64 \pm 0.13$ \\
& 129rr & $9.05 \pm 1.02$ & $15.69 \pm 0.12$ \\
& 129qqrr & $9.15 \pm 0.93$ & $15.68 \pm 0.11$ \\
& 129ss & $6.60 \pm 0.92$ & $16.94 \pm 0.15$ \\
& 129tt & $4.08 \pm 0.87$ & $16.56 \pm 0.23$ \\
& 129sstt & $5.36 \pm 0.72$ & $16.26 \pm 0.15$ \\
2014 \os & 131ss & $5.78 \pm 1.01$ & $16.18 \pm 0.19$ \\
& 131tt & $5.39 \pm 0.99$ & $16.26 \pm 0.20$ \\
& 131sstt & $5.77 \pm 0.81$ & $16.18 \pm 0.15$ \\
& 131qq & $\leq 2.4$ & $\geq 17.13$ \\
& 131rr & $\leq 2.1$ & $\geq 17.28$ \\
& 131qqrr & $\leq 1.5$ & $\geq 17.57$ \\
2014 \pn & 132v & $2.67 \pm 1.22$ & $17.02 \pm 0.50$ \\
& 132w & $3.72 \pm 1.19$ & $16.66 \pm 0.35$ \\
& 132vw & $3.19 \pm 0.86$ & $16.83 \pm 0.29$ \\
& 132x & $2.54 \pm 0.97$ & $17.07 \pm 0.41$ \\
& 132y & $2.15 \pm 0.91$ & $17.25 \pm 0.46$ \\
& 132xy & $2.49 \pm 0.71$ & $17.09 \pm 0.31$ \\
\enddata
\tablecomments{``Target'' is the IAU KBO designation.
``REQID'' is one of the FITS keywords for LORRI images used as a target identifier,
as previously explained.
In addition to presenting data from composite images for a single \mbox{REQID}, 
we also include data from composite images produced by combining all the images
from two REQIDs  that were taken back-to-back on the same day.
\os\ was not detected in REQIDs 131qq and 131rr, and we give 3$\sigma$
upper limits in those cases.
``Signal'' is the total measured signal within a 5-pixel radius aperture centered on the target
and its 1$\sigma$ uncertainty.
We adopted an aperture correction term of $0.10$ to convert from flux within a 5-pixel radius
to flux within an infinite aperture.
``$V$'' is the target's $V$-mag and its uncertainty, assuming the target
has a spectral energy distribution similar to that of KBO Arrokoth.
(\hk\ is an SD, not a CC like Arrokoth, so the color correction in this case might not be
as accurate as for the other objects.)
}
\end{deluxetable*}

\clearpage

\section{Searches for Binaries} \label{sec:binaries}
\subsection{KBO Composite Images} \label{sec:images}
To achieve the sensitivity required to detect the targeted KBOs in LORRI \onebyone format,
we created composite images by combining the 125 images taken for each REQID.
We also created composites by combining all 250 images for the two REQIDs taken
on the same day to improve the SNR even further and produce the deepest imaging
possible for each KBO on a given date.

The fastest apparent (non-sidereal) motion of the KBOs was 
\mbox{$\sim$0\farcs2 s$^{-1}$,}
which means the KBO moved by \mbox{$\leq 0.1$ pixel} relative
to the stars in a single 500~ms exposure.
Thus, the KBO-to-star relative motion is negligible for a single LORRI \onebyone image.
However, the KBOs moved by up to $\sim$26~pixels relative
to the stars over the full time duration needed to obtain the 125 images in a single REQID.
This KBO-to-star relative motion must be taken into account when
producing composite images.
The pointing drift within the ACS deadband described earlier
must also be taken into account when producing composite images.

For this program we produced two types of composite images:
one set co-aligned on the KBOs (using the KBO ephemeris to account for the 
KBO-to-star apparent motion; this smears the stars in the field)
and another set co-aligned on the stars detected in the images (this smears the KBOs).
Before co-adding the images, we first remove any residual background light by
subtracting linear fits to the background level, first row-by-row then column-by-column.
The former removes systematic sub-DN level horizontal structure in the LORRI images, while the latter
removes the ``jailbars'' pattern sometimes seen in LORRI images
\citep[cf.,][]{weaver:2020pasp}.
We use the World Coordinate System (WCS) keywords,
which were calculated using astrometric solutions with the Gaia catalog stars
detected in each image, to determine the appropriate shifts to apply when combining all the images.
To mitigate the effects of pixel outliers (e.g., pixels affected by cosmic rays),
we employ a robust averaging process to produce the composite images.
Briefly, the pixel values for a stack at a particular (x,y) CCD location
are compared to their median, and pixels are rejected if their values
differ by more than $\pm$$3\sigma$ from their median,
where $\sigma$ is computed from a noise model for the LORRI CCD,
which includes the Poisson noise from detected photons, CCD read noise,
and noise in the CCD flat field.
Note that this technique also suppresses the brightness of the smeared stars in the target-aligned stacks.
We compare the image of the KBO in its target-aligned composite to the
images of stars in the star-aligned composite to determine whether there is
evidence for KBO binarity.

Here we present composite images that show both 
\mbox{$512 \times 512$} pixel regions centered on the windowed
portion of the CCD and \mbox{$64 \times 64$} pixel regions centered on the KBO 
and a nearby star of comparable brightness.
Careful examination of these composite images (the grey-scale figures displayed
below for all five KBOs) can be used to determine
the quality of the KBO detections and also provide a qualitative assessment of 
the evidence for KBO binarity.

A more quantitative assessment of KBO binarity was performed by fitting PSFs to the data.
An independent image stacking technique similar to, but different from, 
the method described above was used to create double-sampled 
composite images for each of the REQIDs.\footnote{See \citet{porter:submitted} for 
the details on this technique as applied to the \fourbyfour images;
essentially the same technique is applied here for the \onebyone images.}
An empirical LORRI PSF derived from multiple images of calibration star fields, 
and which has remained stable over the entire course of 
the \nh mission \citep{weaver:2020pasp}, was used to perform an initial
fit to the composite KBO images.
For each stack of 125 images, we fit a model PSF by allowing it to shift in 
(X,Y), and multiplied by a scaling factor,
to minimize a $\chi^2$ estimate as the sum of the square of the 
difference between the model and real images.
If the KBO were single, the difference image should look approximately like random noise.
Binary KBO images would show systematic deviations in the difference images that
are larger than expected from random noise.
As discussed further below, the composite images of \jy\ and the composite image 
from the first epoch of \os\  were not fit very well with single PSF solutions.
Indeed, single PSF solutions for both \jy\ and \os\ 
(Fig. \ref{fig:DoublePSF-JY31} and Fig. \ref{fig:DoublePSF-OS393}) 
produced residual images that appeared
to be over-subtracted in the center and under-subtracted at two points just outside the center.
This is a clear signature of the KBOs appearing as two blended PSFs
with just enough separation to be  resolved.

Recognizing these deficiencies in the single-PSF solutions, we repeated the estimation procedure, 
but with two PSFs in the model rather than one.
The two PSFs were required to be within 20 re-projected pixels ($\approx$10\arcsec) of each other and have 
positive flux values.
For the KBOs that appeared to be single (especially \hk\ and \hz), the two-PSF solutions 
favored keeping the two PSFs on top of each other, and then splitting flux between them.
These solutions thus had effectively the same $\chi^2$ for both single and double PSFs 
(see Fig. \ref{fig:DoublePSF-HK103} and Fig. \ref{fig:DoublePSF-HZ102}).
However, the double PSF solutions for \jy\ and for the bright epoch of \os\ were best fit with two
well-separated PSFs of similar flux 
(Fig. \ref{fig:DoublePSF-JY31} and Fig. \ref{fig:DoublePSF-OS393}).
For \jy\ in particular, the differences between the single and double PSF residuals in 
Figure~\ref{fig:DoublePSF-JY31} are striking.
The differences between the residuals for one and two PSFs are less noticeable for \os, 
as it has both a lower SNR than \jy\ (5 vs 20) and is elongated in the same direction 
as the LORRI PSF.
As discussed further below, \os\ is not visible at all in the second epoch because it was 
near the dimmest point of its rotational lightcurve.
The case of \pn\ is generally similar to those \hk\ and \hz, but \pn's lower SNR 
(3 vs 10) makes discrimination between the single-PSF and double-PSF problematic.

Both single PSF and double PSF fits to the composite images were attempted
for each of the KBOs, which are discussed in turn below.
In each case, we show \mbox{$20 \times 20$} double-sampled images centered
on the KBO (i.e.,  \mbox{$10 \times 10$} native pixels) and use false-color
intensity maps to bring out features more clearly.
\subsection{SD 2011 \hk} \label{sec:hk}
\begin{figure}[h!]
\includegraphics[keepaspectratio,width=\linewidth]{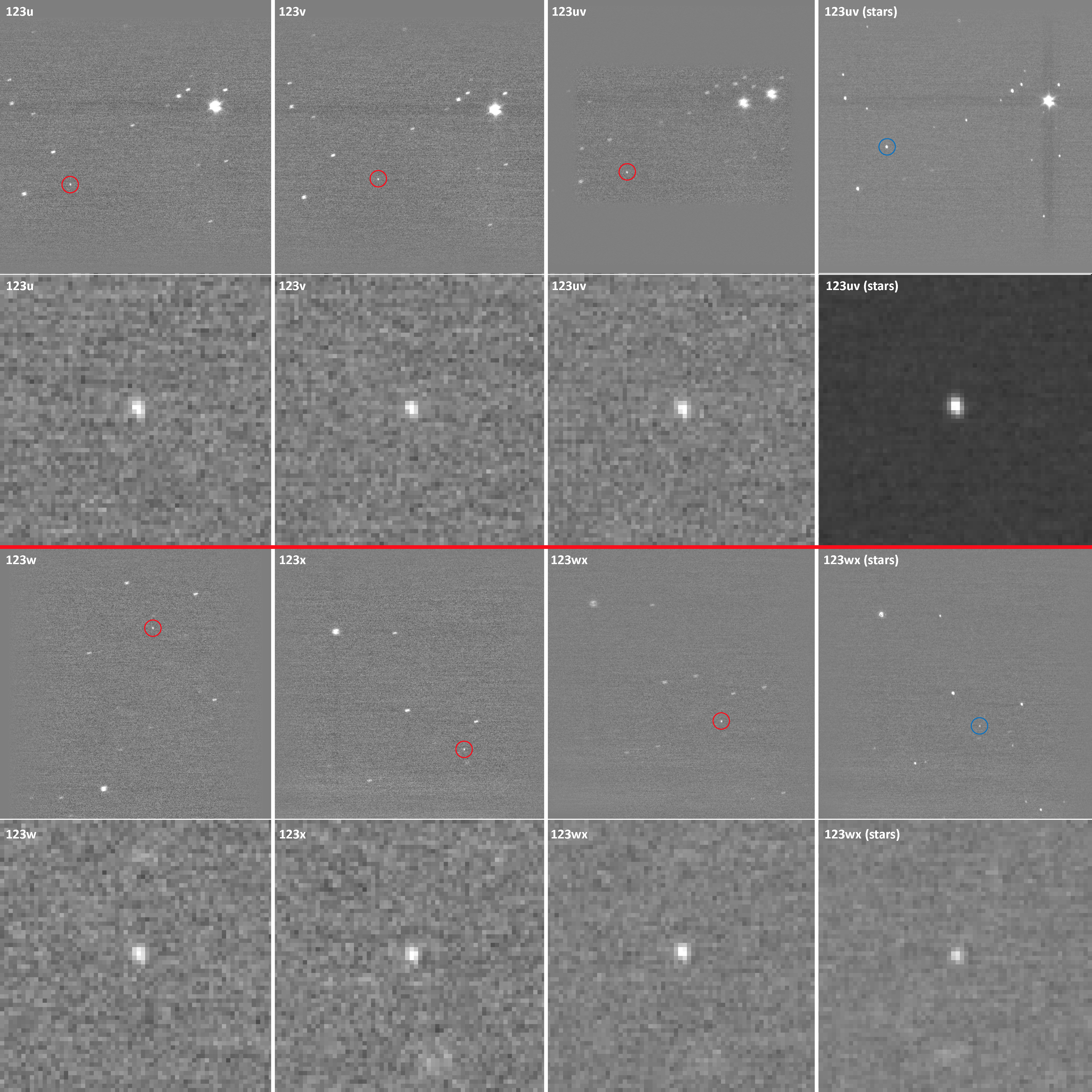}
\caption{Composite images of \hk\ from two different epochs (2018-08-17 above the red line and 
2018-08-19 below the red line) are displayed.
For each epoch, the top row shows \mbox{$512 \times 512$} frames containing the windowed portions of
the full \mbox{$1024 \times 1024$} LORRI \onebyone images, and the bottom row shows
\mbox{$64 \times 64$} frames centered on either the KBO (circled in red in the top row) 
or a nearby star of comparable brightness (circled in blue in the top row).
Composite images produced by co-aligning the images on the ephemeris locations of \hk\ are displayed in the
first three columns: 125 images are combined for the first two columns and 
250 images are combined for the third column.
The REQIDs are displayed in the upper left of each frame.
Composite images created by co-aligning on stars for all 250 images are displayed in the fourth column.
All images are displayed using a linear stretch ranging from $-$1 to 1~DN,
except for the top two images in the fourth column, which use a stretch 
from $-2$ to 4~DN to avoid saturating the circled star.
Celestial north points down and east points to the right in all images. 
}
\label{fig:hk}
\end{figure}

\begin{figure}
    \plottwo{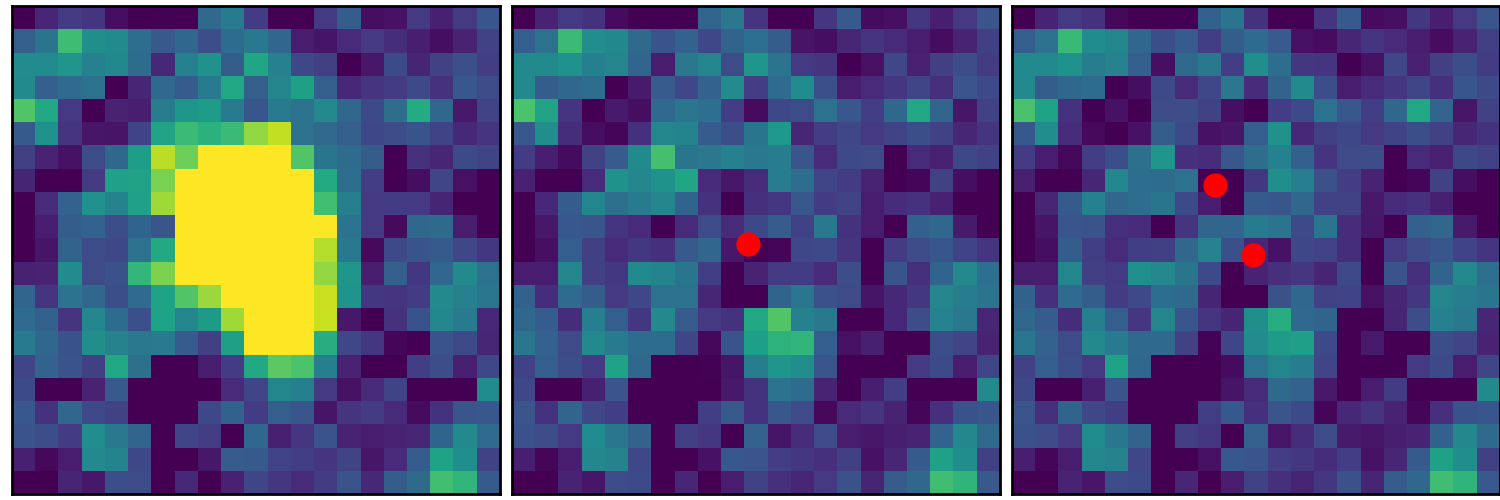}
           {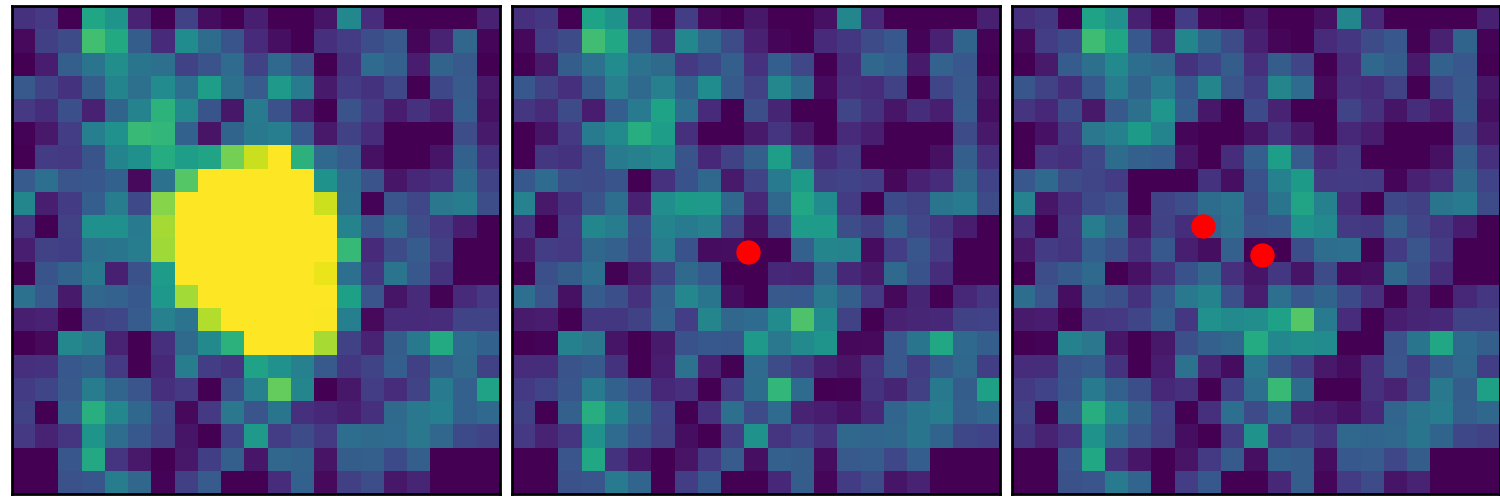}\\
    \plottwo{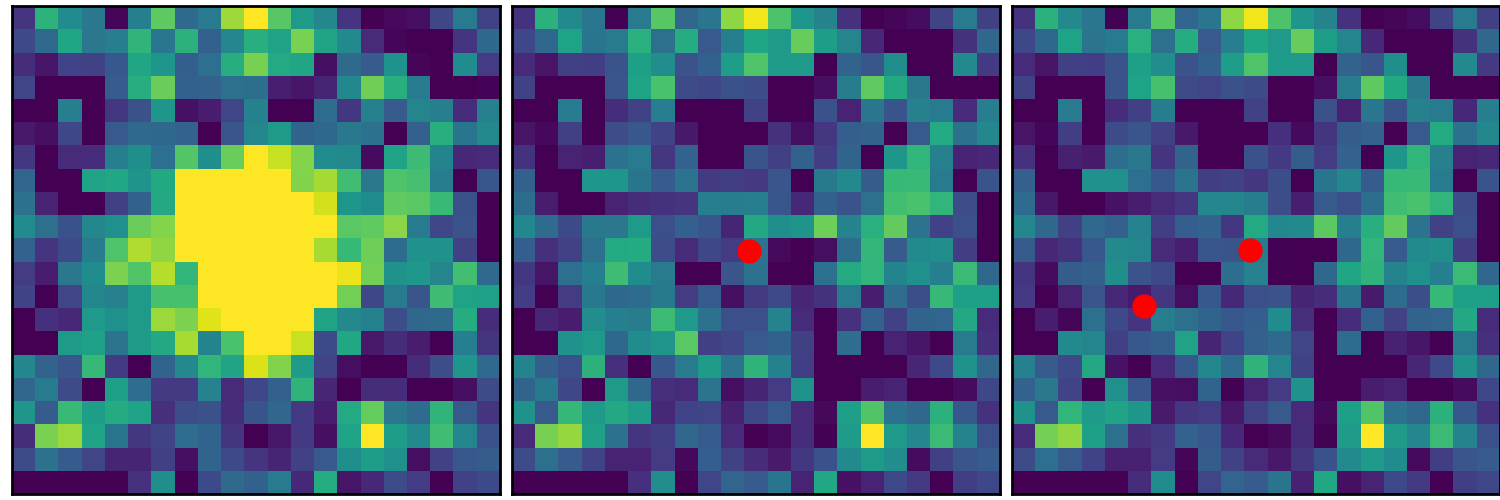}
           {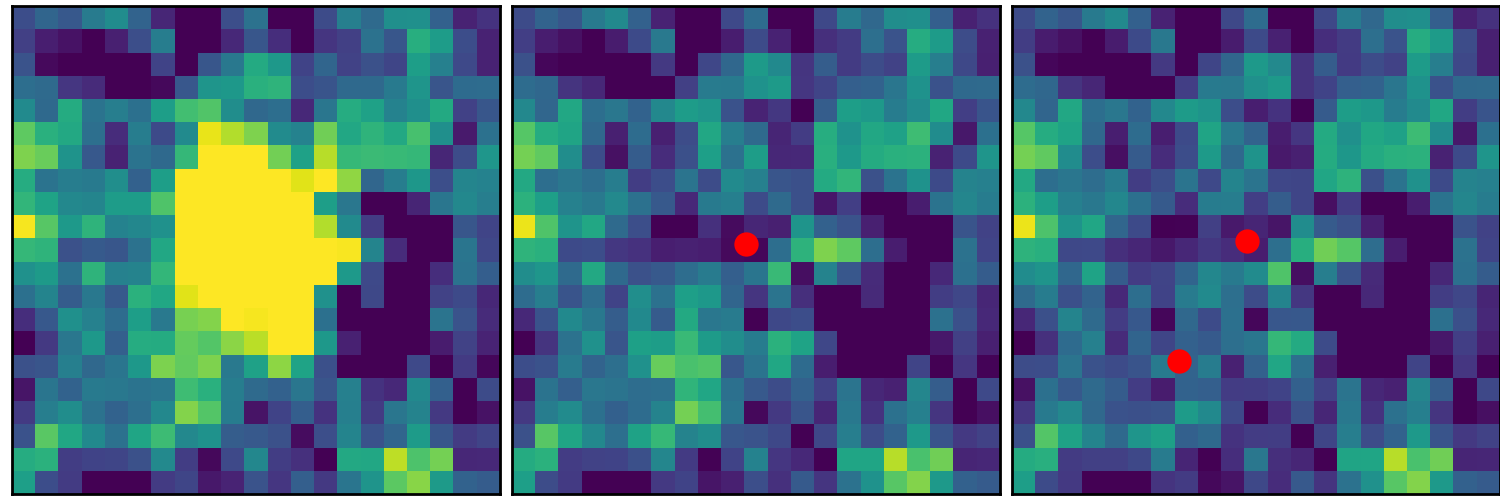}
 \caption{Single and double PSF fits to the double-sampled LORRI images of \hk. 
 The pair of panels on the left hand side shows 
 the two observations from the first epoch (REQIDs 123u and 123v), and the pair of panels on 
 the right hand side shows the two observations from the second epoch (REQIDs 123w and 123x). 
For the three frames within each observation epoch, 
the leftmost image shows the WCS-driven shift stack of the 125 images, the middle image shows the 
residuals after subtracting a best-fit single PSF, 
and the rightmost images show the residuals after subtracting the best-fit double-PSFs. 
The red dots show the  locations of the center of each PSF fit. 
The same linear stretch is used for each image. 
Since the residuals in the double PSF solution look similar to those in the 
single PSF solution, and the locations of the object locations in the double PSF solution
varies over the image, \hk\ does not appear to be a binary.
}
\label{fig:DoublePSF-HK103}
\end{figure}

SD KBO \hk\ is relatively bright and could be easily detected in each of the 
four composite images (Fig.~\ref{fig:hk}).
The composite images of \hk\ are slightly elongated, but so are the images
of nearby stars and the PSF fitting procedure (Fig.~\ref{fig:DoublePSF-HK103}) 
does not provide convincing evidence for binarity.
Thus, we do not claim that \hk\ is binary based on the \nh \onebyone data.
\hst observations of \hk\ also did not show any direct evidence of binarity \citep{benecchi:hst-dkbos},
albeit at $\sim$10$\times$ poorer spatial resolution than provided by the \nh program.
The light curves of \hk\ at multiple solar phase angles derived from LORRI \fourbyfour
data \citep{verbiscer:2019dkbos} reveal a relatively short rotational period (10.83012~h) with
an amplitude that varies from 0.2~mag at a phase angle of 51$\arcdeg$ to 0.9~mag at
a phase angle of 96$\arcdeg$
and do not show any evidence of binarity.
\clearpage

\subsection{CC 2011 \jy} \label{sec:jy}
\begin{figure}[h!]
\includegraphics[keepaspectratio,width=\linewidth]{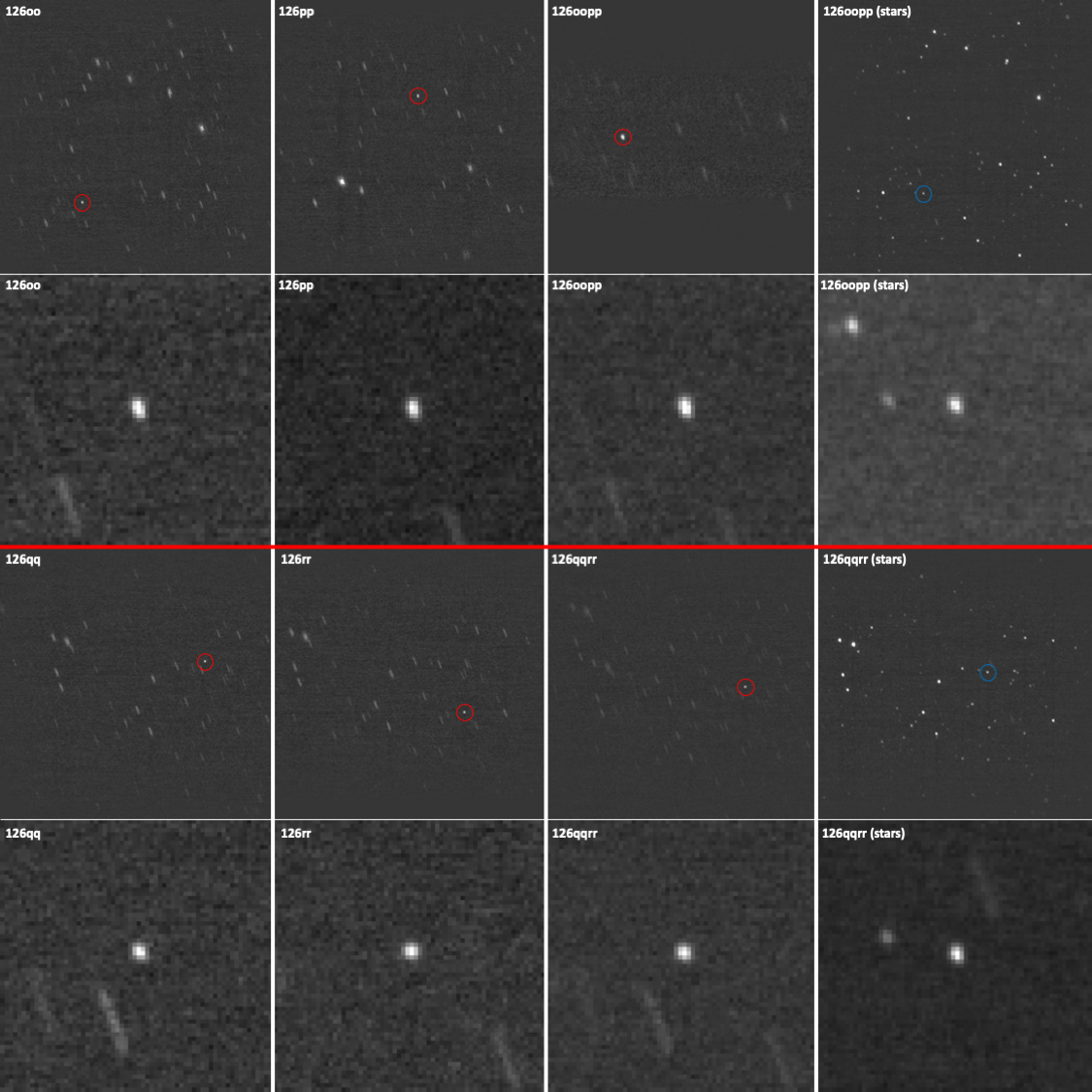}
\caption{Composite images of \jy\ from two different epochs (2018-09-05 above the red line and 
2018-09-11 below the red line) are displayed.
The layout is exactly as previously described for \hk.
In the first epoch of images co-aligned on the KBO's ephemeris, \jy\ is clearly elongated 
in a diagonal direction relative to the stellar images.
In the second epoch of images co-aligned on the KBO's ephemeris, \jy\ appears to be
broader horizontally relative to the stellar images.
Thus, there is evidence for binarity for both epochs.
All images are displayed using a linear stretch ranging from $-$1 to 2.5~DN,
except \mbox{$64 \times 64$} 126pp and \mbox{$64 \times 64$} 126qqrr (stars) are 
displayed on a linear scale from $-$1 to 3~DN to avoid saturating on the brightest
pixel and \mbox{$64 \times 64$} 126oopp (stars) is displayed on a linear scale
from $-1$ to 2~DN to bring out the fainter stars.
Celestial north points down and east points to the right in all images. 
}
\label{fig:jy}
\end{figure}

\begin{figure}
    \plottwo{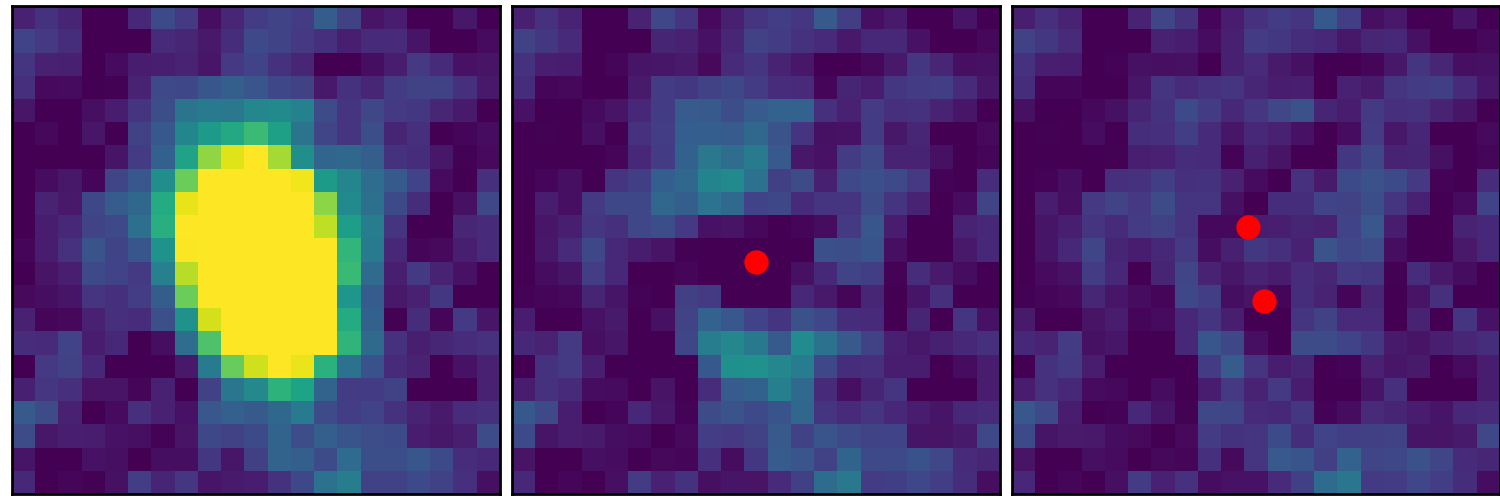}
            {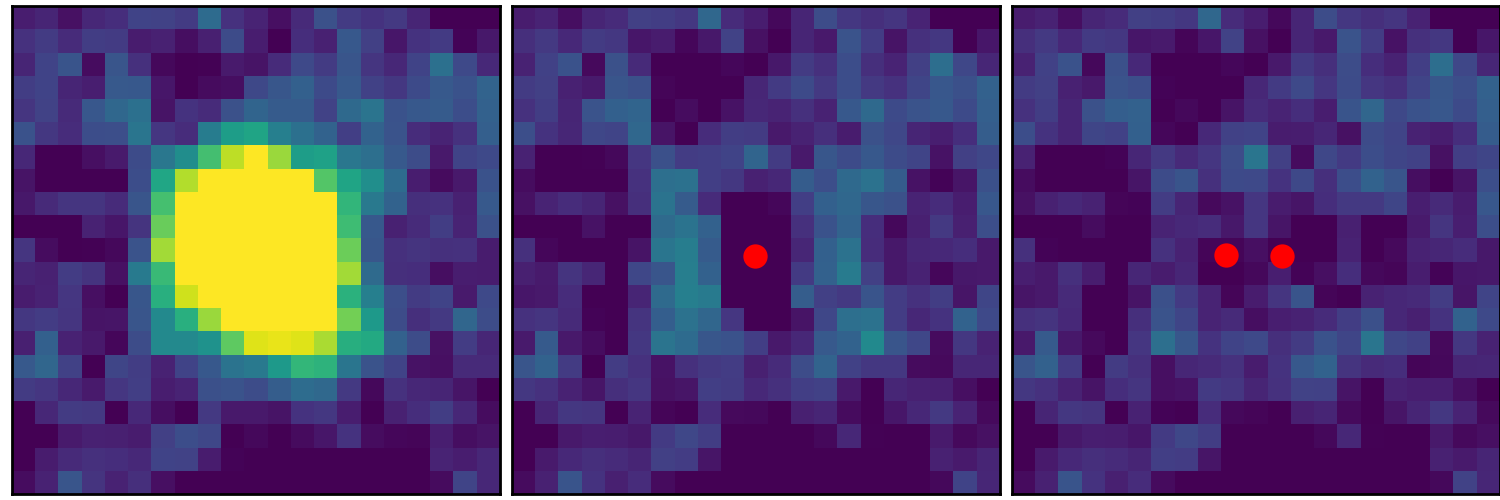}\\
    \plottwo{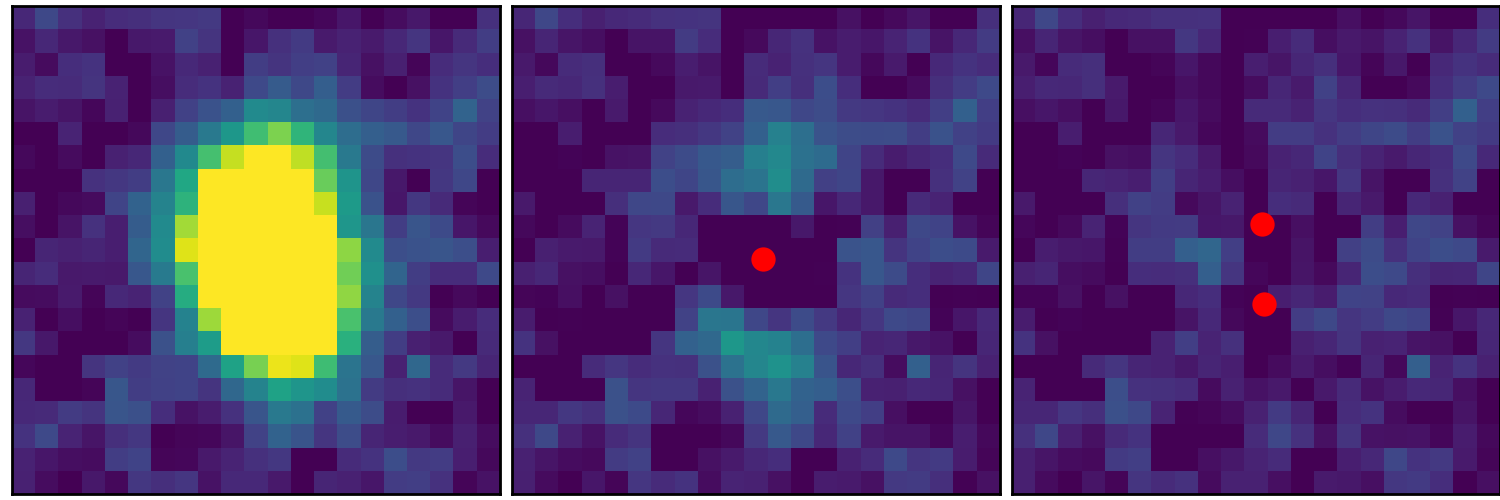}
            {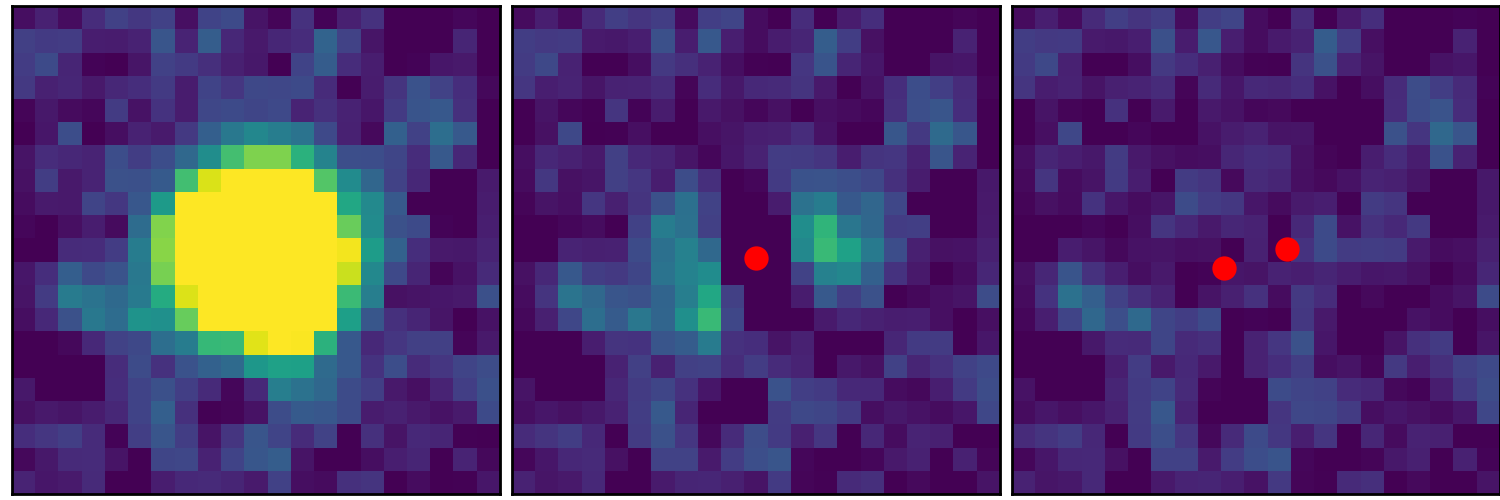}
\caption{Single and double PSF fits to the double-sampled LORRI images of CC \jy. 
The pair of panels on the left hand side shows 
the two observations from the first epoch
(REQIDs 126oo and 126pp), and the pair of panels on the right hand side shows the two observations from the 
second epoch (REQIDs 126qq and 126pp). 
See Figure \ref{fig:DoublePSF-HK103} for detailed descriptions of each panel. 
Here, the double PSF solution is clearly much better than  the single PSF solution, 
and the location of the two PSFs is consistent between the two observations in each epoch. 
The ``high-low-high'' residuals in the single PSF fits are typical of two barely resolved PSFs.
We thus conclude that \jy\ is almost certainly a binary.
}
\label{fig:DoublePSF-JY31}
\end{figure}

CC KBO \jy\ is also relatively bright and was easily detected in each 
of the four composite images (Fig.~\ref{fig:jy}).
Close inspection of these images shows that \jy's appearance is broader
than the PSF in all four cases.
For both of the first two composite images (from REQIDs 126oo and 126pp),
 \jy\ appears to be elongated along a diagonal direction.
For both of the second two composite images (from REQIDs 126qq and 126rr),
which were taken $\sim$34~h after the first pair, \jy\ appears to be
extended in the horizontal dimension relative to the PSF.
These impressions are reinforced by the PSF fitting analysis (Fig.~\ref{fig:DoublePSF-JY31}),
which shows that the double PSF solution is clearly better than the single PSF solution.
As discussed further in Section~\ref{sec:jy-binary},
these morphological features in the images of \jy\
can be associated with the motion of two bodies in a compact binary system.
\clearpage
\subsection{CC 2011 \hz} \label{sec:hz}
\begin{figure}[h!]
\includegraphics[keepaspectratio,width=\linewidth]{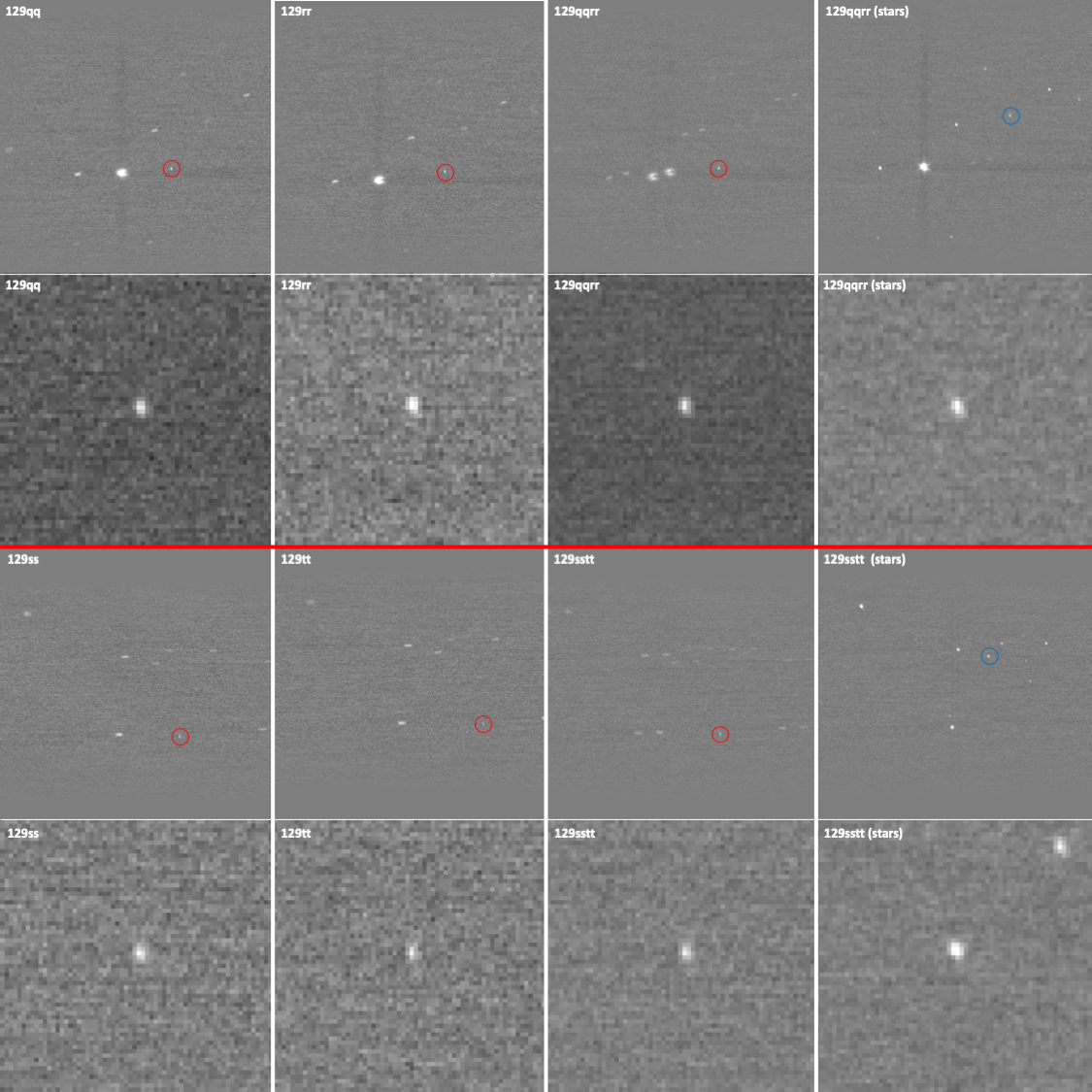}
\caption{Composite images of \hz\ from two different epochs (2018-11-30 above the red line and 
2018-12-04 below the red line) are displayed.
The layout is exactly as previously described for \hk.
In all the composite images, \hz\ has a similar appearance to the PSF (i.e., to the stars),
which shows there is no evidence for binarity in either epoch.
All images are displayed using a linear stretch ranging from $-$1 to 1~DN,
except \mbox{$64 \times 64$} 129qq and \mbox{$64 \times 64$} 129qqrr are 
displayed on a linear scale from $-$1 to 1.5~DN to avoid saturating on the brightest
pixel.
Celestial north points down and east points to the right in all images. 
}
\label{fig:hz}
\end{figure}

\begin{figure}[h!]
    \plottwo{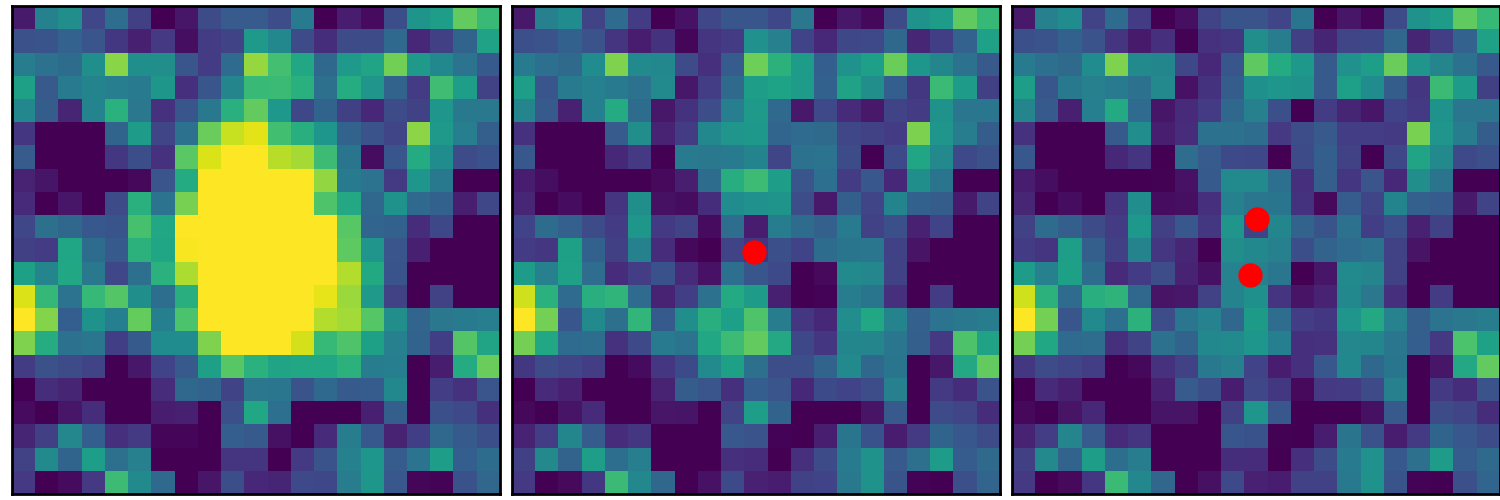}
            {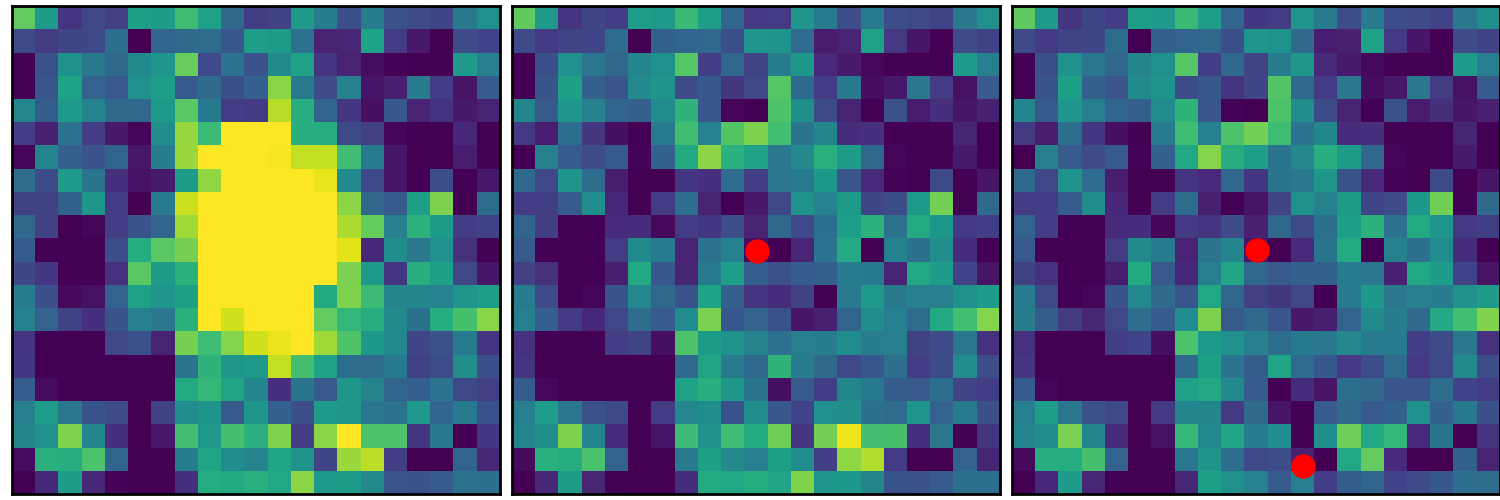}\\
    \plottwo{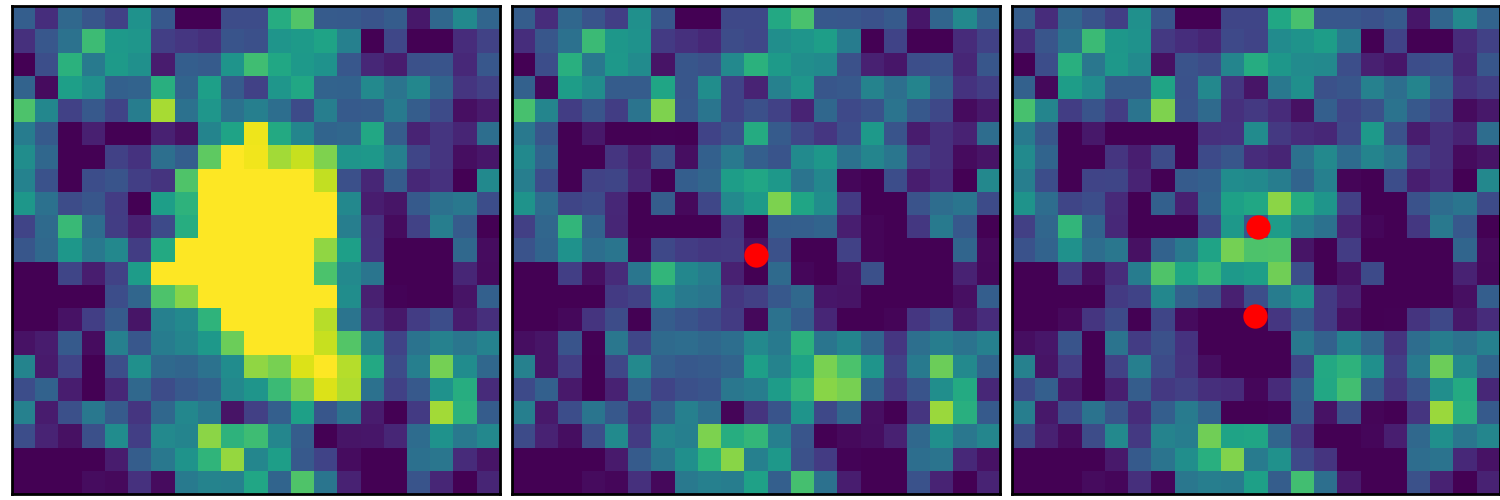}
            {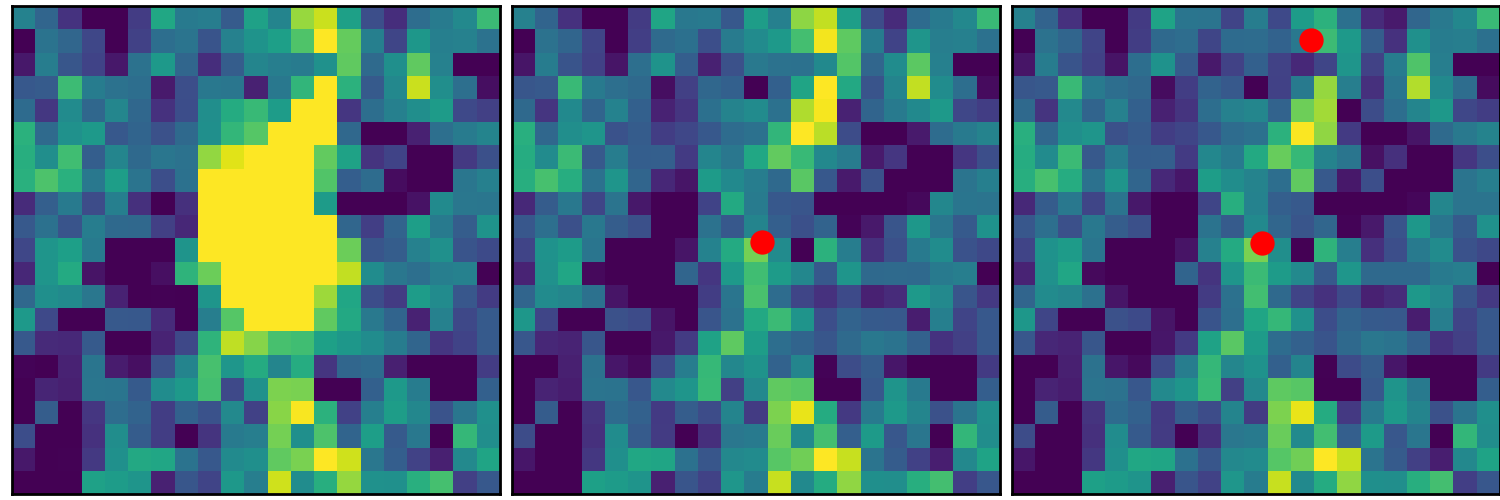}
    \caption{Single and double PSF fits to the double-sampled LORRI images of \hz.
The pair of panels on the left hand side shows the two observations from the first epoch
(REQIDs 129qq and 129rr), and the pair of panels on the right hand side shows the two observations from the 
second epoch (REQIDs 129ss and 129tt). 
See Figure \ref{fig:DoublePSF-HK103} for detailed descriptions of each panel. 
Much like \hk, \hz\ is well-fit with a single PSF. 
While there is a consistency in the angle of the two PSFs in the first epoch, the dimmer second
PSF flips sides relative to the primary PSF, and so is likely not real.
}
\label{fig:DoublePSF-HZ102}
\end{figure}

CC KBO \hz\ has intermediate brightness and was clearly detected in each 
of the four composite images (Fig.~\ref{fig:hz}).
Close inspection of these images shows that \hz's appearance is stellar-like in all four cases,
which suggests that \hz\ is not a resolved binary system.
This conclusion is supported by the more sophisticated PSF fitting analysis (Fig.\ref{fig:DoublePSF-HZ102}),
which shows that the single and double PSF solutions have similar residuals. 
\clearpage
\subsection{CC 2014 \os} \label{sec:os}
\begin{figure}[h!]
\includegraphics[keepaspectratio,width=\linewidth]{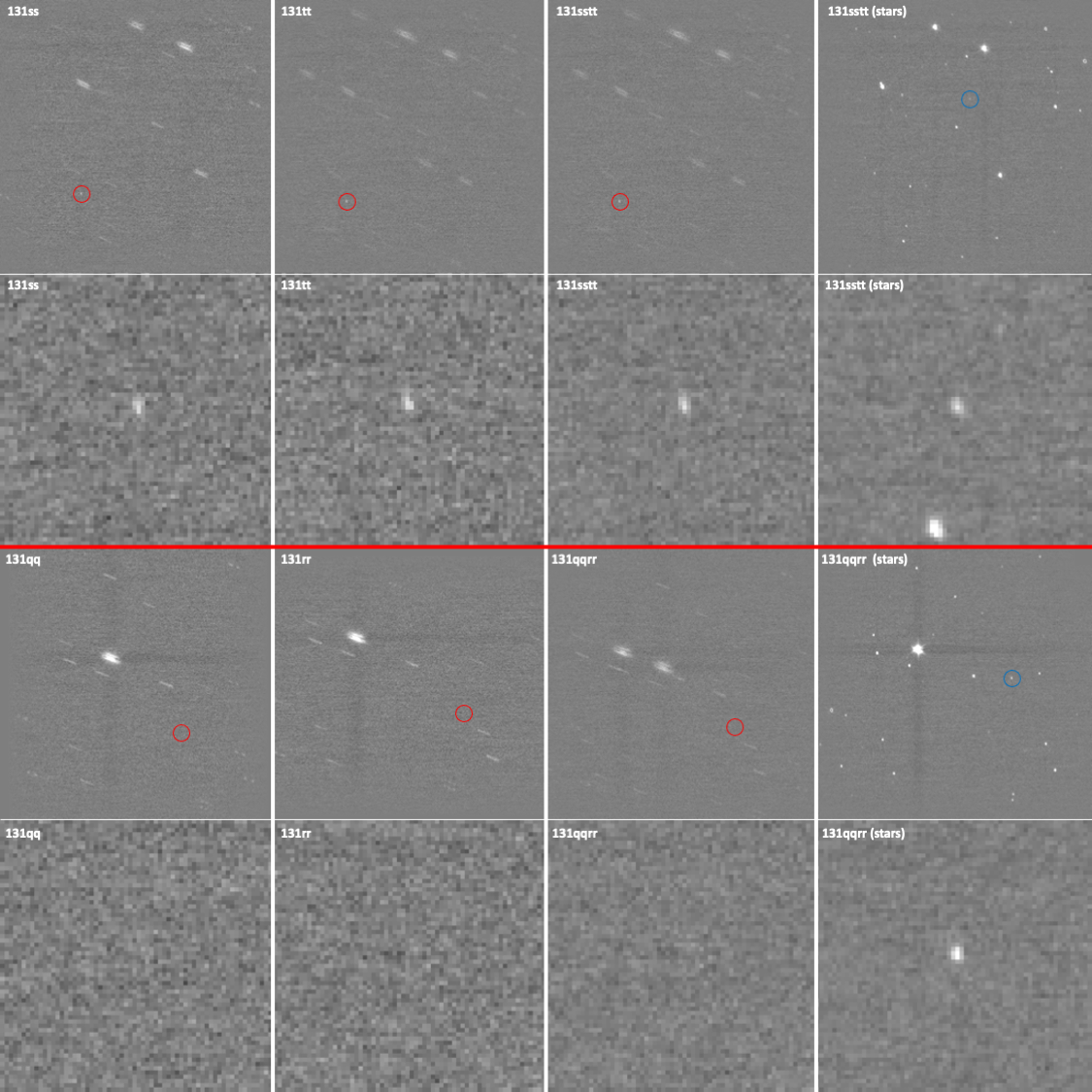}
\caption{Composite images of \os\ from two different epochs (2019-01-05 above the red line and 
2019-01-06 below the red line) are displayed.
The layout is exactly as previously described for \hk.
\os\ appears to be elongated relative to the expected PSF in the first epoch suggesting
it is a binary object.
Unfortunately, \os\ could not be detected in the second epoch when it was near the
minimum in its lightcurve brightness.
All images are displayed using a linear stretch ranging from $-$1 to 1~DN.
Celestial north points down and east points to the right in all images. 
}
\label{fig:os}
\end{figure}

In the first set of images of CC KBO \os, the object is detected with relatively poor SNR,
and it is not detected at all in the second set (Fig.~\ref{fig:os}).
We calculate SNR values of 5.7, 5.4, and 7.1 for REQIDs 131ss, 131tt, 
and their combination, respectively.

\begin{figure}[h!]
\includegraphics[keepaspectratio,width=\linewidth]{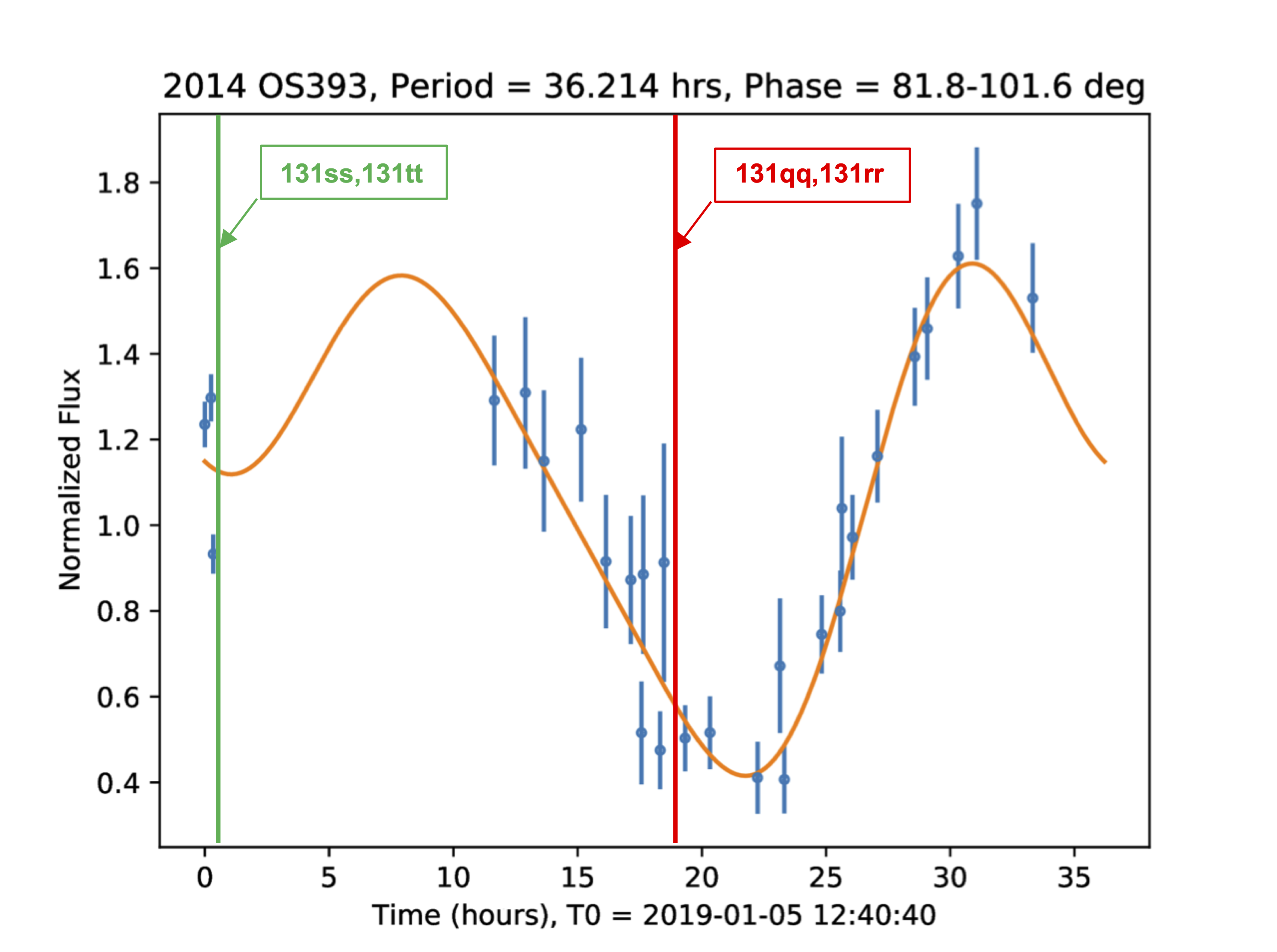}
\caption{\os\ lightcurve data from \citet{porter:submitted} are plotted showing 
the strong variation of signal over time.
The observation time for REQIDs 131ss and 131tt (green vertical line) occurred near
the time of maximum brightness, whereas REQIDs 131qq and 131rr (red vertical line)
were conducted near the time of minimum brightness.
}
\label{fig:os-lightcurve}
\end{figure}

In the first set of images, \os\ appears to be more extended than for 
any of the other KBOs discussed here and is more extended than a 
similarly bright nearby star, providing evidence that \os\ is probably a binary system.
If this interpretation is correct, the satellite is $\sim$150~km \mbox{(2.24~pixels)} from the primary at 
celestial position angle of $\sim$207\arcdeg. 
\clearpage
Unfortunately, \os\ wasn't detected in the second epoch (REQIDs 131qq and 131rr),
probably because it was near its lightcurve minimum (see Fig.~\ref{fig:os-lightcurve}).
We know that we searched in the correct location because the
ephemeris for \os\ is accurate to $\sim$1~pixel.
Based on the photometry from the LORRI \fourbyfour program \citep{porter:submitted},
we estimate that \os\ was about two times fainter for the second epoch compared
to the first, which pushed the object below our detection limit.
%

\begin{figure}[h!]
    \plottwo{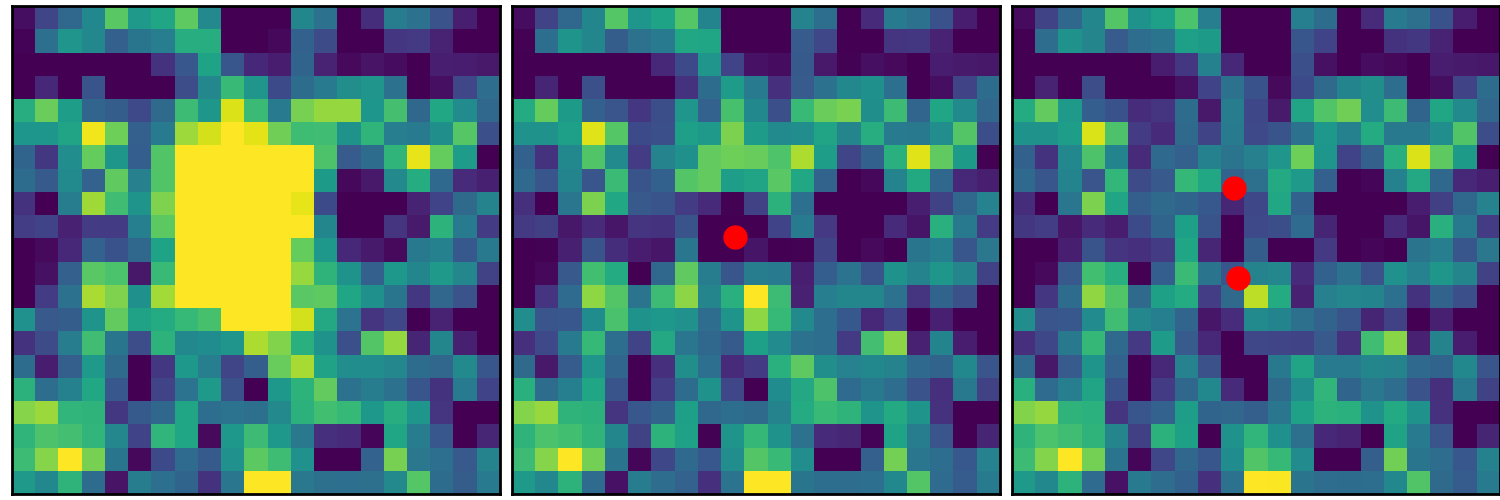}
            {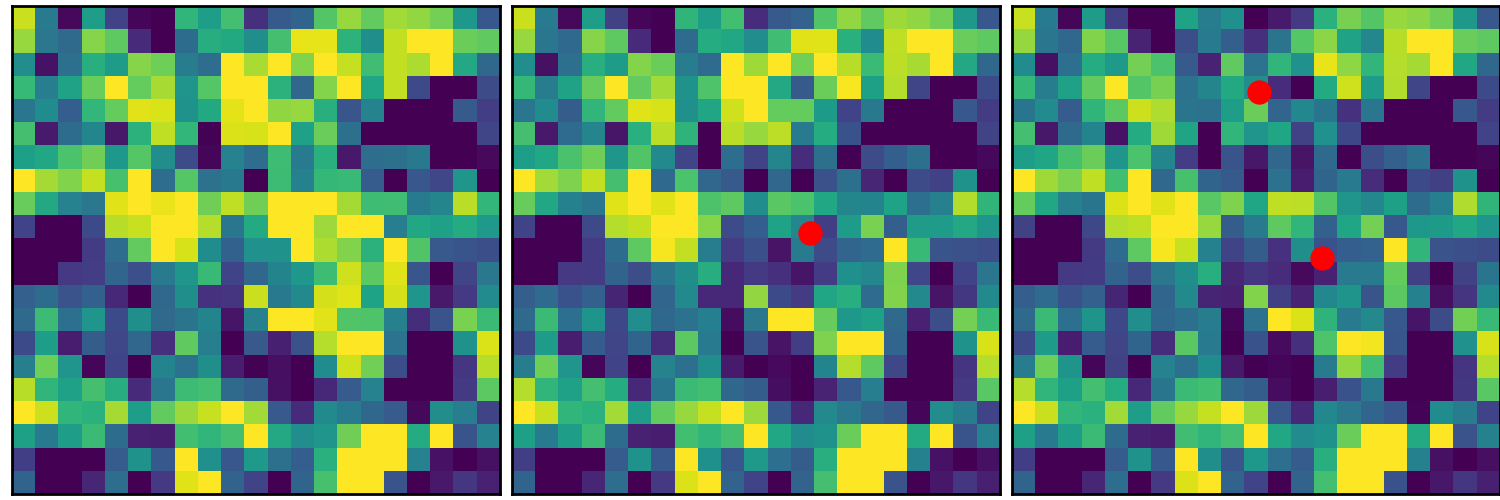}\\
    \plottwo{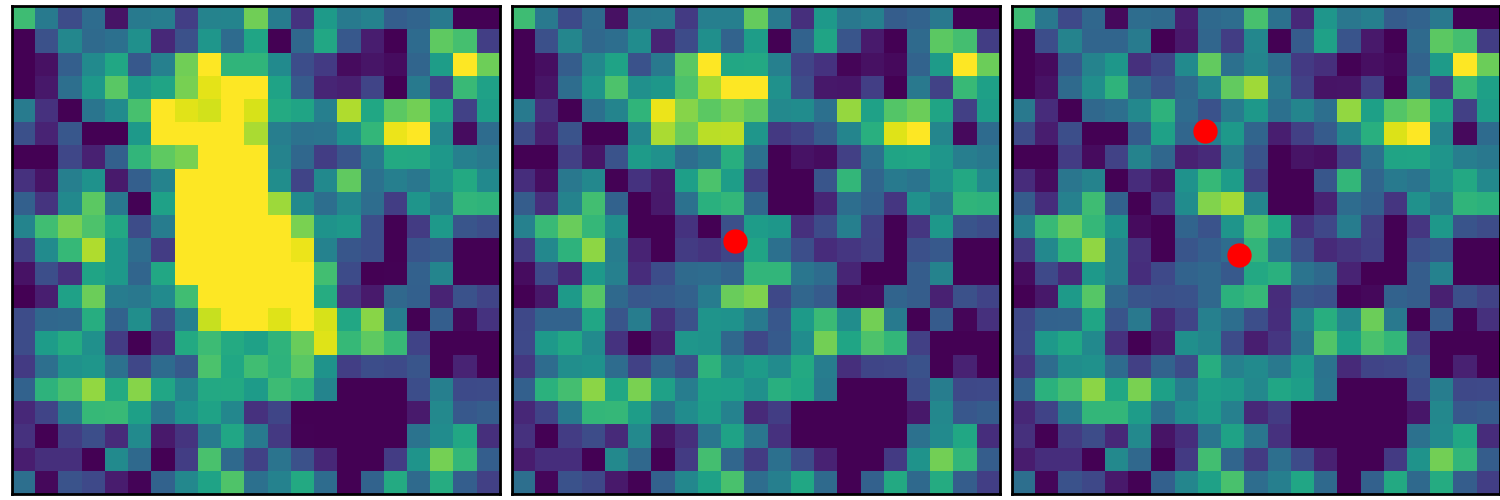}
            {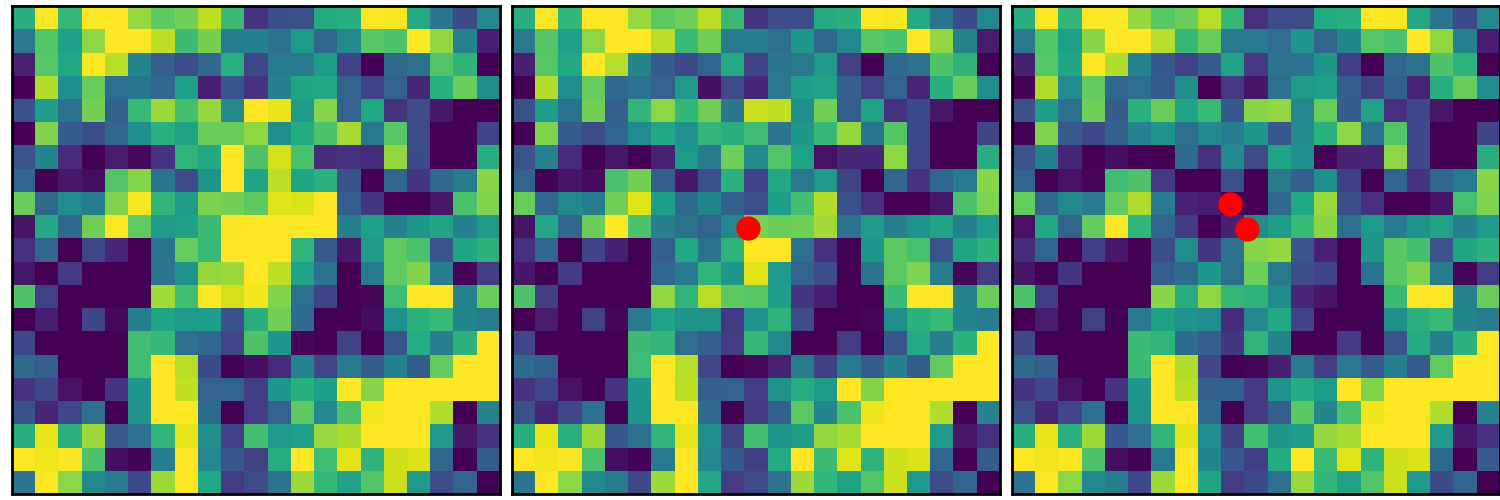}
    \caption{Single and double PSF fits to the double-sampled LORRI images of \os.
The pair of panels on the left hand side shows the two observations from the first epoch
(REQIDs 131ss and 131tt), and the pair of panels on the right hand side shows the 
two observations from the second epoch (REQIDs 131qq and 131rr). 
See Figure \ref{fig:DoublePSF-HK103} for detailed descriptions of each panel. 
The images of \os\ in the first epoch appear to be elongated.
Moreover, the locations of the PSFs in the double PSF solutions are
roughly consistent in the two separate image sets taken in the first epoch.
However, \os\ is not visible at all in the second epoch.
This is consistent with its lightcurve (see Fig.~\ref{fig:os-lightcurve}), 
which predicts that the KBO should be about one magnitude dimmer in the 
second epoch than the first.
}
\label{fig:DoublePSF-OS393}
\end{figure}

Our PSF-fitting analysis of \os's images also suggests that this KBO 
is a binary (Fig.~\ref{fig:DoublePSF-OS393}), but the low SNR in the
first set of observations and the lack of a clear detection of the object in the
second set make this conclusion more tentative than for \jy.

At Earth's distance from \os\ during opposition ($\sim$42~au),
the binary separation (if real) is only $\sim$0\farcs005, 
which means imaging detection of the binary nature is unlikely with currently available facilities.
\os\ was discovered by \hst but no direct evidence for binarity was seen in those images,
which had a resolution of \mbox{$\sim$2400 km}
(2 pixels, at  $0\farcs04$ per pixel).
We note that the relatively long lightcurve period ($\sim$36~h) is consistent with
\os\ being a tidally locked binary system.
Perhaps precision lightcurve photometry, or occultation measurements \citep{porter:submitted}, can shed
further light on the possible binary nature of \os.
\subsection{CC 2014 \pn} \label{sec:pn}
\begin{figure}[h!]
\includegraphics[keepaspectratio,width=\linewidth]{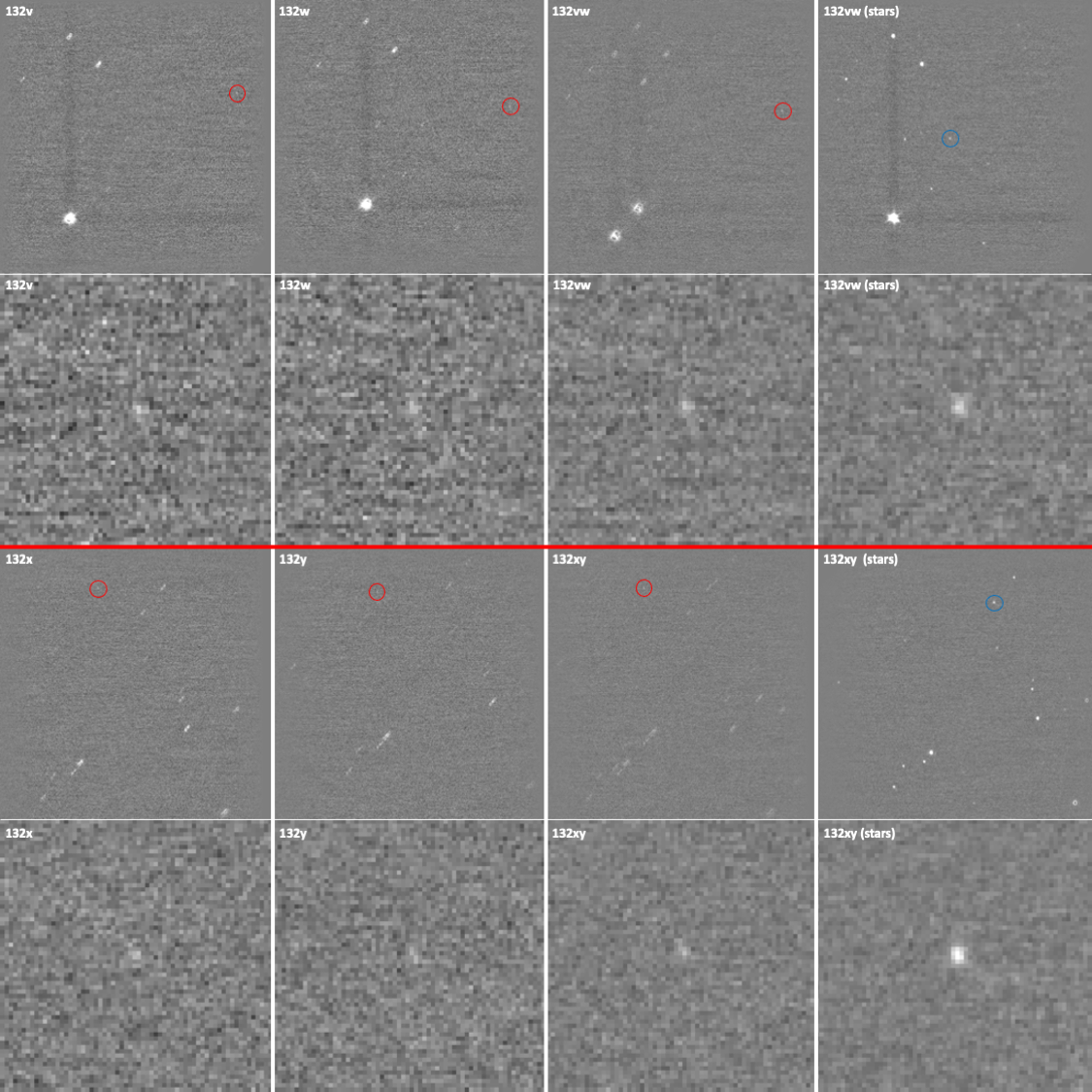}
\caption{Composite images of \pn\ from two different epochs (2019-03-09 above the red line and 
2019-03-11 below the red line) are displayed.
The layout is exactly as previously described for \hk.
\pn\ is detected in all these images but with low SNR.
All images are displayed using a linear stretch ranging from $-$1 to 1~DN.
Celestial north points down and east points to the right in all images. 
}
\label{fig:pn}
\end{figure}

Unlike all the other cases considered here, CC \pn\ was not always in the windowed region of the CCD.
For REQIDs 132v, 132w, 132x, and 132y, \pn\ was in 55, 52, 76, and 88 images, respectively, out of
the total of 125 images in each case.
As previously mentioned, the \pn\ background field was relatively sparse, which made it problematic to
design a window that contained both the KBO and at least one star brighter than $V=13$.
A compromise we made was to put the single bright star near one edge of the window and
the KBO near another edge.
Unfortunately, the KBO moved in and out of the windowed region as the LORRI boresight drifted 
within the specified deadband.
We used the predicted ephemeris of \pn\ to find all the images when the KBO was within the window
and also at least 5 pixels from the edge, and we created composites using 
only these ``good'' images (Fig.~\ref{fig:pn}).

\begin{figure}[h!]
    \plottwo{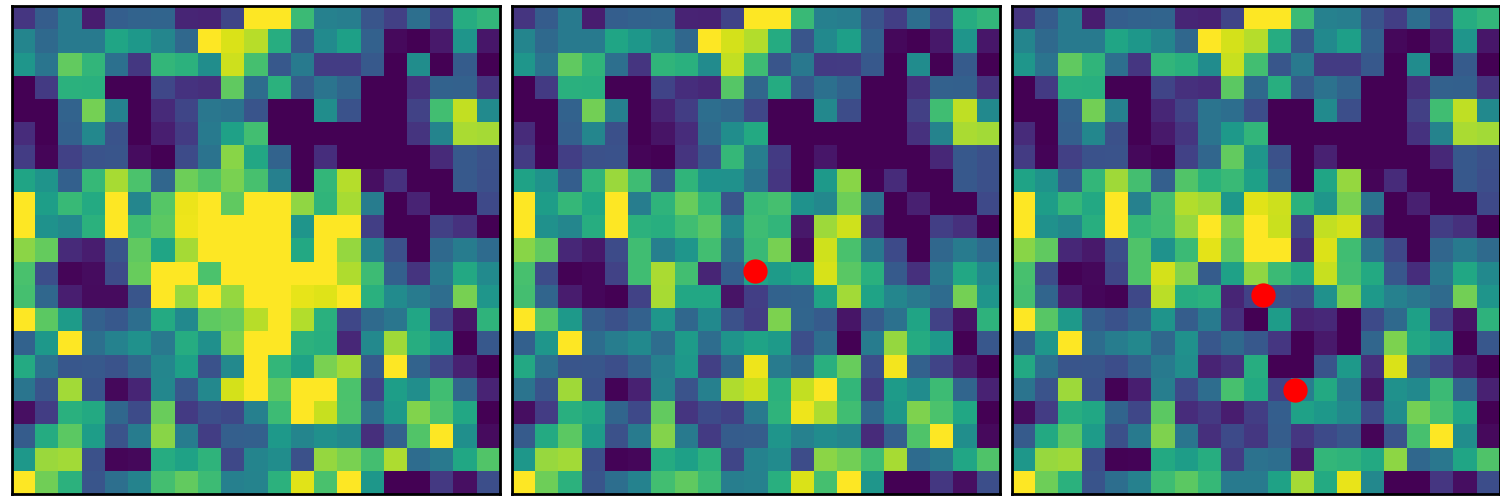}
            {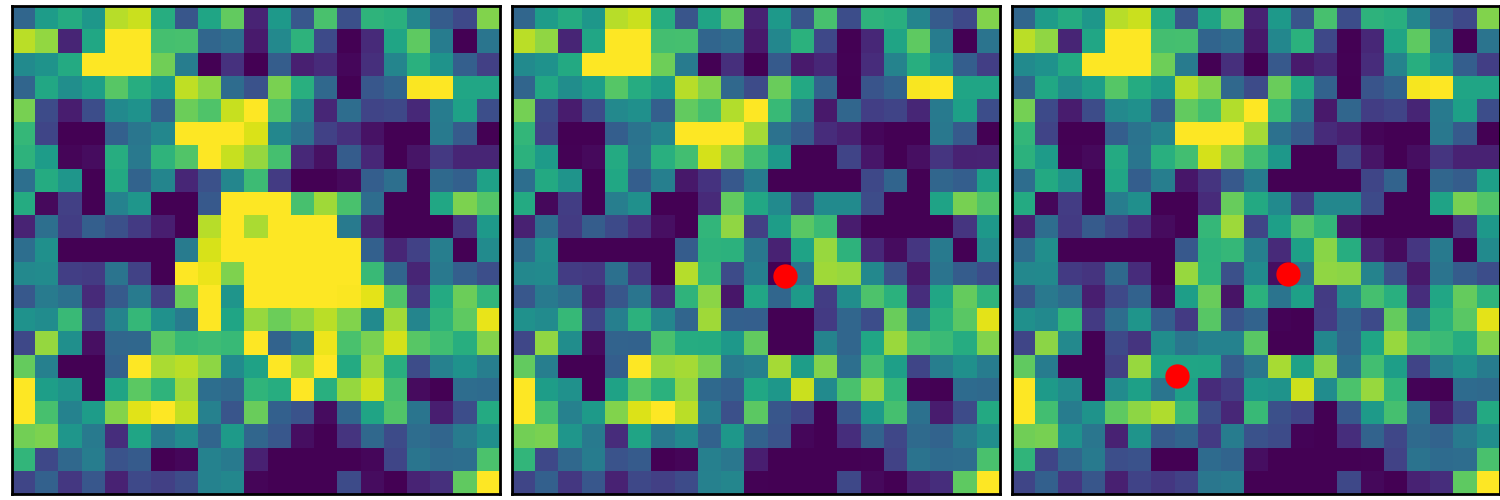}\\
    \plottwo{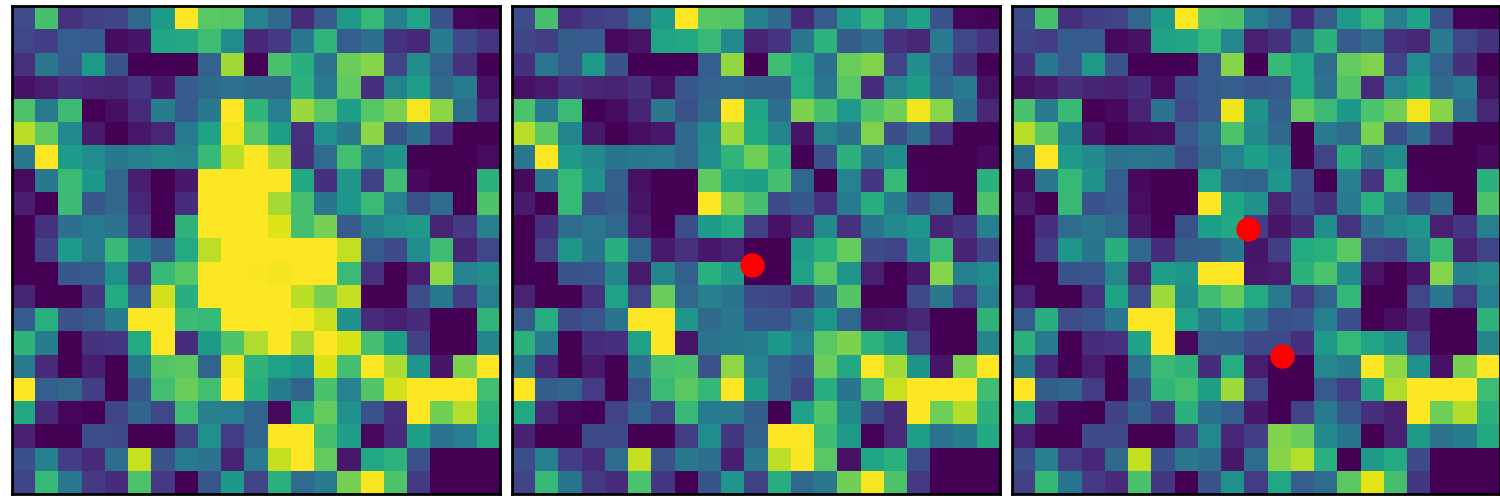}
            {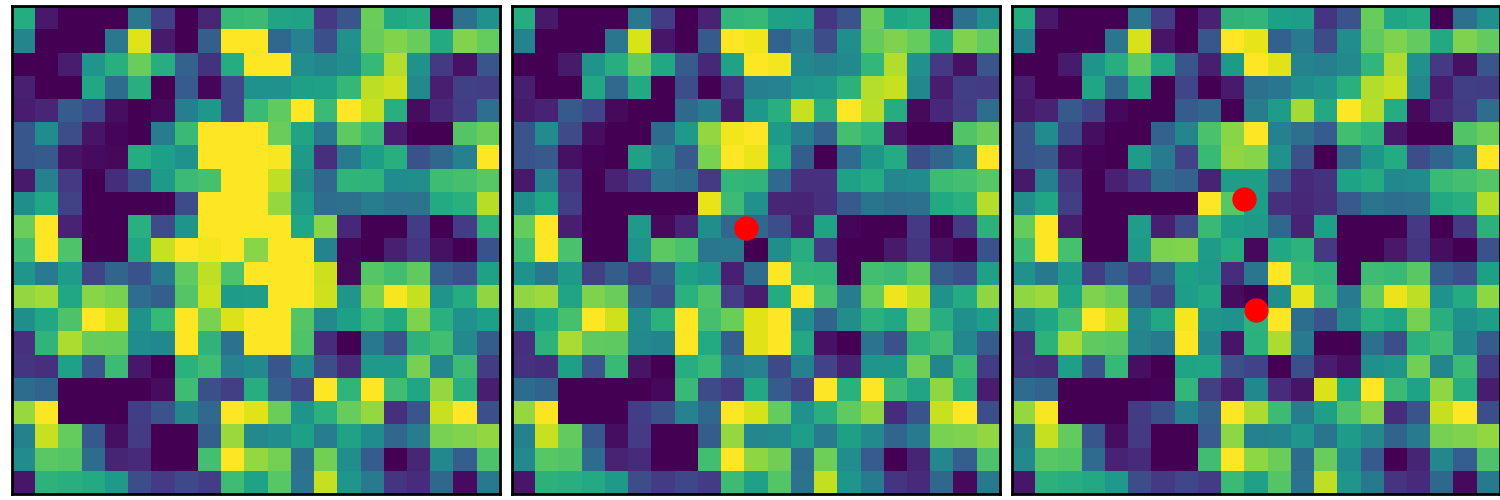}
    \caption{LORRI 1x1 observations for \pn. The pair of panels on the left hand side shows 
    the two observations from the first epoch
    (REQIDs 132v and 132w), and the pair of panels on the right hand side shows the two observations from the 
    second epoch (REQIDs 132x and 132u). See Figure 
    \ref{fig:DoublePSF-HK103} for detailed description. While \pn\ was detected in all four observations, 
    it had a low SNRs in both epochs. In none of the observations was the two-PSF 
    solution significantly better than the single PSF solution, which suggests there is
    no evidence that \pn\ is binary.
    }
    \label{fig:DoublePSF-PN70}
\end{figure}

 \pn\ was also exceptionally faint, essentially at the sensitivity limit of LORRI.
 The SNRs were 2.2, 3.1, and 3.7 for REQIDs 132v, 132w, and 132vw, respectively.
 (The image of \pn\ from 132w had an anomalously narrow PSF, which suggests it
 may have been affected by a weak cosmic ray event or a pixel with larger than
 typical dark current.)
 The SNRs were 2.6, 2.4, and 3.5 for REQIDs 132x, 132y, and 132xy, respectively.
 For all these latter cases, there are pixels of comparable brightness to the peak located
 near the ephemeris location of \pn\ that make photometry in this case especially difficult.
 The photometry (Table~\ref{tab:photometry}) gives $V \approx 17$ for these latter
 observations.
 Given the poor SNR for the \pn\ observations, we cannot draw any definitive conclusions
 about its binarity.
 The same conclusion is reached by the more refined analysis (Fig.~\ref{fig:DoublePSF-PN70}).
\clearpage

\subsection{Binary Orbit Determination of \jy} \label{sec:jy-binary}
While \jy\ appears to be binary in both epochs, those observations alone are not sufficient
to determine the orbital elements of the binary system.
Each LORRI observation provides two data points, ($\delta$RA and $\delta$Dec), which means two observations
can only provide four constraints.
While we were able to extract a distinct astrometric point for each 125-image stack, for a total of four 
observations and eight constraints, they were close enough to each other ($\sim$2\% of the orbit)
that they were not effectively independent points.
The high SNR of \jy\ in the LORRI \onebyone images meant that the uncertainty in the relative astrometry of the
secondary with respect to the primary was small ($\sim$0\farcs04), or about
2-3\% of the 1\farcs2-1\farcs7 apparent separation of the two bodies.
A bound Keplerian orbit with an unknown system mass has seven free parameters, which can be described as 
the relative position/velocity state vector of the second with respect to the primary, plus the total system
mass.
Alternatively, the orbit can be parameterized as the system semimajor axis, eccentricity, 
longitude of the ascending node, argument of the periapse, anomaly at epoch, and orbital period.
If we constrain the mutual orbit to be circular, the eccentricity of the orbit is fixed at zero, and the 
argument of the periapse becomes undefined (as there is no discrete periapse).
The number of free parameters for a circular orbit with unknown mass is therefore five,
but a unique solution is still not possible with only two astrometric observations.
While binary KBOs have been detected at a large range of 
mutual eccentricities \citep{grundy:kbo-binaries-2019}, 
all binary KBOs with a semimajor axis smaller than 3000 km have been found to have circular mutual orbits
\citep{grundy:kbo-binaries-2019}, with the exception of (42355) Typhon-Echidna, which may have been
perturbed by encounters with Neptune \citep{grundy:typhon2008}.
This result implies that these tight KBO orbits have been circularized by mutual body tides raised on the two 
objects \citep{porter:binaries2012}.
Since any plausible mutual orbits for \jy\ are smaller than this, we can reasonably assume
that \jy's binary has a circular orbit.

\begin{figure}[h!]
\includegraphics[keepaspectratio,width=\linewidth]{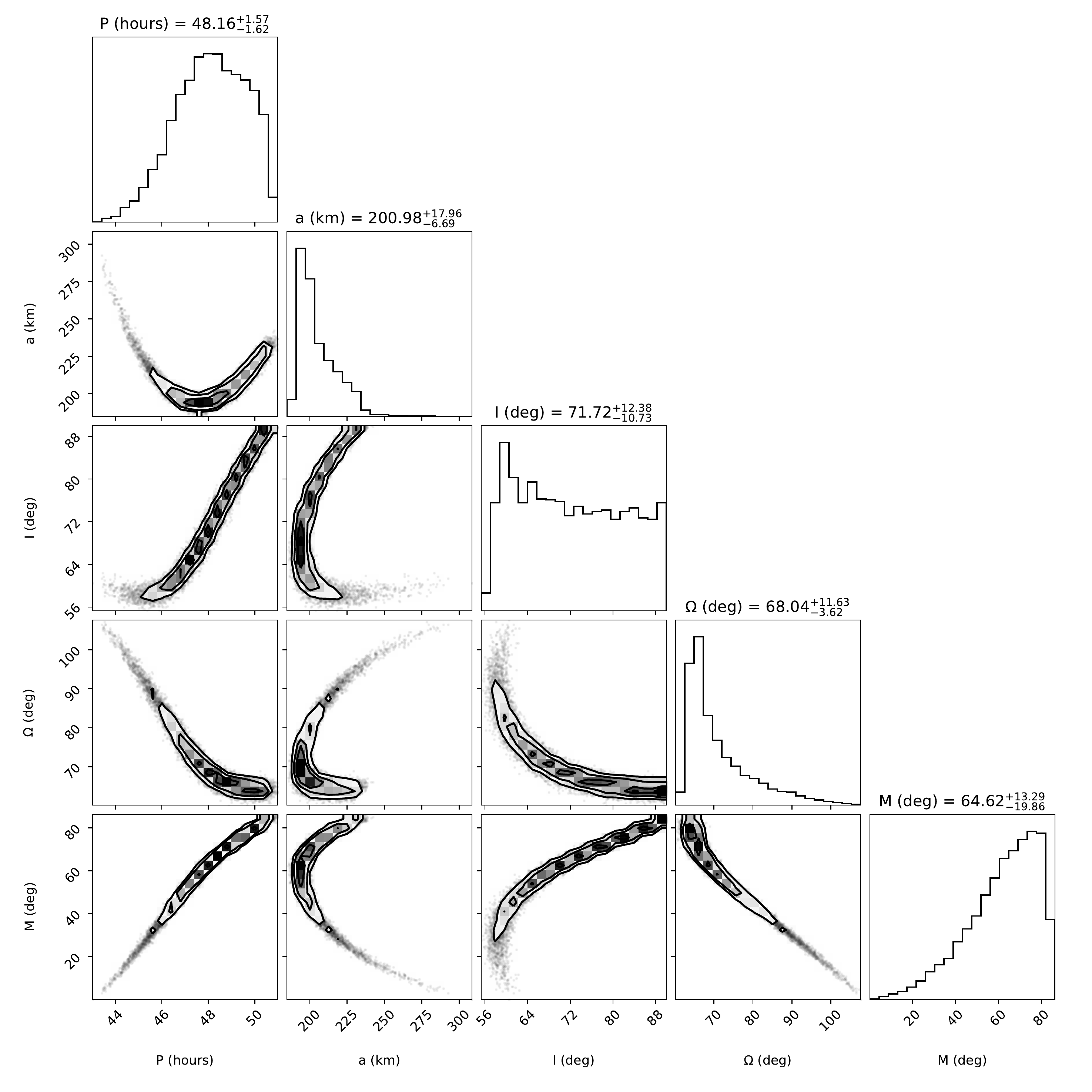}
\caption{The probability distribution function (PDF) of the mutual orbit solution for \jy's binary
orbit using relative astrometry from the LORRI \onebyone images (see Fig.~\ref{fig:DoublePSF-JY31}). 
The orbit is assumed to be circular, and the fit parameters are period, orbital radius, 
inclination, longitude of the ascending node, and mean anomaly at the epoch. 
For this solution, the period is unconstrained, which results in a wide spread in the fit parameters.}
\label{fig:BinOrbitA}
\end{figure}

To test how constraining the eight-but-really-four parameters were on the circular solution space, we used 
\textit{emcee} \citep{foreman:mcmc-2013,foreman:mcmc-2019} to create a 
Markov chain Monte Carlo (MCMC) solution cloud, shown in Figure \ref{fig:BinOrbitA}.
There were a range of solutions that were consistent with a roughly two day orbit,
but the solutions were highly-correlated in semimajor axis and period, and the orbit pole was essentially
unconstrained.
The LORRI \onebyone resolved observations alone could therefore not be used to fit the mutual orbit of \jy.


\begin{figure}[h!]
\includegraphics[keepaspectratio,width=\linewidth]{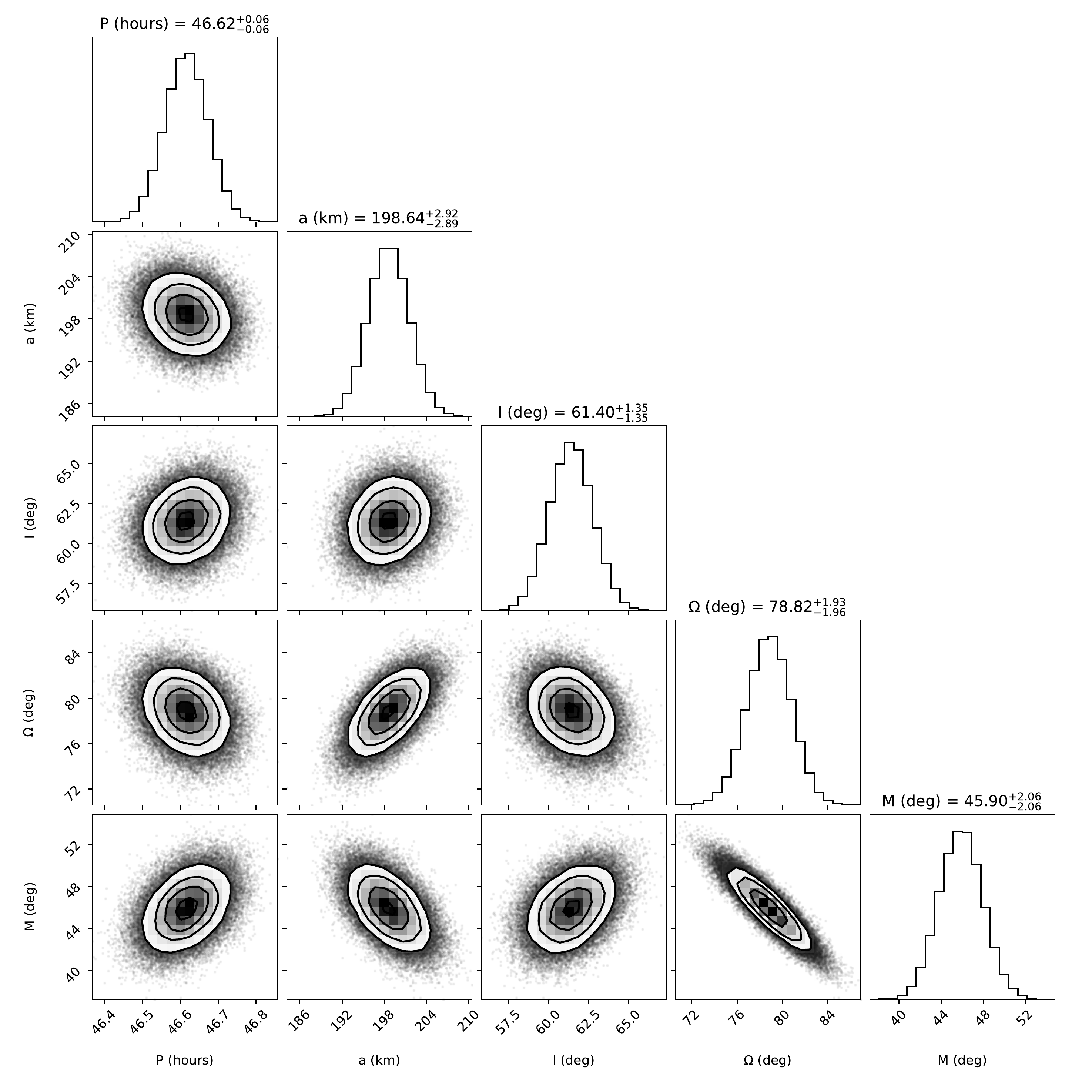}
\caption{The same as for Fig.~\ref{fig:BinOrbitA}, except for this solution the period is
constrained by the lightcurve results from the \fourbyfour data \citep{porter:submitted},
which results in much tighter distributions of the fit parameters.}
\label{fig:BinOrbitB}
\end{figure}

In addition to the LORRI \onebyone images, \nh also observed \jy\ in \fourbyfour format
at five different epochs, with 18 observations spanning $\sim$36~h in each epoch, 
including observations just before and just after the \onebyone observations \citep{verbiscer:2019dkbos}.
These observations allowed us to constrain the 
rotational period of \jy\ to be \mbox{46.6 $\pm$ 0.05 h.}
Because the orbit is so tight, and because we had already assumed that it has tidally circularized,
we made the further assumption that the two bodies of the \jy\ system are tidally locked,
such that their rotation period is equal to their orbit period.
If the orbit is indeed circular, this should be a reasonable assumption because \citet{porter:binaries2012}
found that tidal de-spinning to a double-synchonous configuration was very rapid after similar-sized 
binary KBOs achieved tidal circularization.
The lightcurve itself also suggests tidal synchronization, as only the one period was evident in the 
solution, with no beat frequencies that might be expected from two objects rotating 
at different rates \citep{verbiscer:2019dkbos, porter:submitted}.
This result provided a fifth constraint on the five parameters, allowing us to determine a proper solution set
for the mutual orbit.
As shown in Figure \ref{fig:BinOrbitB}, the combination of the LORRI lightcurve period constraint 
and the LORRI \onebyone 
resolved astrometry was sufficient to provide an excellent constraint on the mutual orbit of \jy.
We were able to constrain the semimajor axis of circular orbit to be \mbox{198.6 $\pm$ 2.9 km} and the inclination of 
the mutual orbit to be \mbox{61$\fdg$34 $\pm$ 1$\fdg$34} to the ICRF pole, 
or  \mbox{61$\fdg$5 $\pm$ 1$\fdg$3} to \jy's heliocentric orbit pole.
While binary KBOs observed from Earth typically have a mirror-ambiguity that leads to two conjugate orbit 
pole solutions \citep{grundy:kbo-binaries-2019}, the \nh LORRI \onebyone observations of \jy\ were 
7$\arcdeg$ apart in solar phase angle.
This, plus the ``sense-of-motion'' detectable within each day of data, was sufficient to break the degeneracy 
and constrain \jy's mutual orbit to only one family of solutions.
The orbital elements of \jy\ derived from our analysis are provided in Table~\ref{tab:jyorbit}.

\begin{deluxetable*}{cccccc}
\tablecaption{Orbital Elements of Binary Cold Classical KBO 2011 \jy \label{tab:jyorbit}}
\tablewidth{0pt}
\tablehead{
\colhead{a} & \colhead{e} & \colhead{i} & \colhead{$\Omega$} & \colhead{M} & \colhead{P} \\
\colhead{(km)} & & \colhead{(deg)} & \colhead{(deg)} & \colhead{(deg)} & \colhead{(hours)}
}
\startdata
$198.6 \pm 2.9$ & 0 & $61.5 \pm 1.3$ & $78.8 \pm 1.9$ & $45.90 \pm 2.06$ & $46.62 \pm 0.06$ \\ 
\enddata
\tablecomments{``a'' is the semi-major axis of the binary orbit,
``e'' is the eccentricity, ``i'' is the inclination relative to the object's heliocentric orbit pole,
``$\Omega$'' is the longitude of the ascending node of the binary orbit, ``M'' is the
mean anomaly at the epoch of observation, and ``P'' is the binary orbital period.
The binary orbit's inclination relative to the ICRF pole is \mbox{61$\fdg$34 $\pm$ 1$\fdg$34}.
}
\end{deluxetable*}

With a physical separation of $\sim$200 km, \jy's components are almost certainly tidally
locked to each other.
From the orbit, the total mass of the system is \mbox{$\sim$$1.7\times10^{17}$ kg.}
If we assume each component has a density of 500 kg/m$^3$, 
similar to comet 67P/Churyumov-Gerasimenko \citep{jorda:cg2016},
and that the two components are roughly equal in size,
then the two components would have equivalent radii of $\sim$34~km, making them at least 5.8 radii apart.
If the bodies have a density of \mbox{1000 kg m$^{-3}$}, 
their equivalent radii would be $\sim$27~km, putting them at least 7.4 radii apart.
Tidal circularization occurs rapidly at separations  less than 10 radii \citep{porter:binaries2012}, 
and tidal locking occurs shortly thereafter.

The equivalent radii derived from the dynamical analysis are significantly larger than the
values estimated from the visible photometry (see Table~\ref{tab:properties}).
The dynamical analysis suggests that \emph{each} component of the \jy\ primary 
has an equivalent spherical diameter of at least 56~km, assuming they are less dense than water ice, 
whereas Table~\ref{tab:properties} suggests that their diameters should be $\sim$42~km,
assuming the geometric albedo is 15\% and the quoted visible magnitude is actually 
the sum of the light from two equally bright objects.
However, \citet{porter:submitted} gives \mbox{$H_{0} = 8.1$} for \jy, which raises the photometrically
derived size up to $\sim$58~km for each component of the binary, fully consistent with the
dynamically derived size.
\section{Discussion} \label{sec:discussion}
In our small survey of KBOs passing close to the \nh spacecraft, we found that
two out of four CCs are binary, while the single SD KBO searched does not appear
to have a companion $\geq$450~km from the main object.
One of the CCs searched (\pn) was too faint to draw definitive conclusions about
its possible binarity, which essentially means that two of three CCs surveyed by \nh have
companions.

The observational status of KBO binaries has been summarized 
in a recent review paper \citep{noll:tno2020}, which provides updates to
an earlier review \citep{noll:tno2008}.
These reviews conclude that the binary fraction of SD KBOs is probably less than$\sim$10\%,
whereas at least 20\% of KBO CCs are binaries.
However, these papers acknowledge that the binary fraction estimates suffer
from various selection effects that could bias the results, and at least make
them highly uncertain.

\citet{fraser:2017, fraser:2017correction} discovered a sub-population of ``blue'' objects among the CCs, and they
argued that essentially {\it all} of those objects must have been born as binaries.
They hypothesized that these blue CC binaries must have formed near $\sim$38-40~au and were
subsequently pushed by gravitational interactions with Neptune into the CC region.
These authors invoked the streaming instability to form rotating pebble clouds within the solar
nebula, which subsequently produced larger planetesimals via gravitational collapse.
According to their simulations, binary planetesimals are the predominant large bodies produced during
this process, together with a large number of smaller single objects (the latter are needed to carry
away the angular momentum of the original pebble cloud).
The CCs surveyed by \nh all seem to be red (like most of the CC KBOs; \citet{benecchi:hst-dkbos}), 
which means the high binary fraction for the blue CCs might not apply to the objects \nh surveyed.

However, a new paper \citep{Nesvorny:psj2021} addresses planetesimal formation in the CC region in detail
and concludes that binary formation is likely to be predominant in this region, 
reinforcing their earlier arguments along the same lines \citep{Nesvorny:nature2019}.
These authors also invoke the streaming instability followed by gravitational collapse of pebble clouds as the
primary planetesimal formation mechanism in the CC region, and they demonstrate that
binaries naturally form for a wide range of pebble cloud angular momentum conditions.
Thus, the high binary fraction for the CC KBOs surveyed by \nh is consistent with current CC KBO formation
scenarios. Furthermore, the \nh results extend the previous surveys to smaller objects 
(\mbox{$H_{0} = 5-8$} for the previous surveys versus \mbox{$H_{0} = 8-10$} for {\it New Horizons})
and considerably smaller binary separations ($\geq$2000~km versus $\sim$200~km).
As shown in \citet{porter:binaries2012}, the combination of Kozai-Lidov cycling 
and tidal friction (also known as KCTF) may have produced a large number of very tight binary KBOs, 
which is also consistent with the \nh detections.

The previous surveys showed that most of the CC binaries have orbits with 
small eccentricities (i.e., nearly circular orbits), nearly equal-size bodies, and
tend to be relatively compact (i.e., with semi-major axes a small fraction of the mutual Hill sphere radius),
which is also true for the two CC binaries discovered by {\it New Horizons}.
The vast majority of CC binaries have prograde mutual orbits (i.e., mutual orbital rotations in the same
direction as their heliocentric orbits), which is also true for \jy, although its orbital inclination
is relatively large \mbox{(61$\fdg$34 $\pm$ 1$\fdg$34).}
However, the \nh search was certainly biased toward equal-sized binaries because
those were the only types that could be detected.
That, and other selection effects (e.g., optical resolution limits), might suggest that the reported 
results on KBO binaries generally provide {\it lower limits} to the true binary fraction.

An independent analysis of the CC population's binary fraction has been offered based on a systematic analysis of 
212 CCs observed by the {\it Hubble Space Telescope} \citep{Parker:dps2020}. 
A careful observational debiasing of this sample taking into account
the CC luminosity function, binary formation models, and binary colors suggests that the intrinsic binary fraction for 
objects larger than 30~km and semi-major axes larger than 3000~km is likely lower: 
13-16\% for   \mbox{$H_{0} < 6.2$} and 1-4\% for  \mbox{$H_{0} > 6.2$}. 
However, this survey is limited to significantly brighter objects than our \nh sample. 
Additional analysis by \citet{Parker:psanh2021} considers what the binary fraction for 
tight binaries (between contact and 0.5--1\% of the Hill radius) might be based on an evaluation 
of doublet cratering projected from the currently known wide binary population and results 
from the simulations of tidal circularization timescales for small KBOs by \citet{porter:binaries2012}. 
Parker concludes that while the calculation is sensitive to the proposed separation distribution input, 
there is a high likelihood for a large population of tight binaries, which is consistent with our 
identification of \jy\ and\os\ as binaries. 
Future occultation observations, or observations by \nh of other small KBOs, 
may provide additional support for this conclusion. 

\section{Summary} \label{sec:summary}
We have taken advantage of the \nh spacecraft's passage through the densest portion
of the cold classical (CC) Kuiper belt to image five objects (4 CCs and 1 scattered
disk KBO) with spatial resolutions of \mbox{136-430 km},
which is approximately 6-18 times better than possible from {\it HST}.
By adding hundreds of images taken by the LORRI camera in its
high resolution mode (\onebyone format), we reached a sensitivity limit of \mbox{$V \approx 17$}
and detected all five KBOs.
Adopting reasonable assumptions, we demonstrated that the CC KBO \jy\ is almost 
certainly a binary system with two
equally bright bodies in a nearly circular orbit with a separation of \mbox{$198.6 \pm 2.9$ km}
and an orbital period of \mbox{$1.940 \pm 0.002$ d.}

We conclude that the CC KBO \os\ is also likely a binary with the bodies separated by $\sim$150~km.
However, the low \mbox{SNR ($\approx$5-7)} for the \os\ detections during the first epoch,
and the lack of detection during the second epoch (likely explained because \os\ was near a 
minimum in its lightcurve, which put the object below LORRI's detection limit), 
lowers our confidence level in claiming that \os\ is a binary system.
Future stellar occultation opportunities \citep{porter:submitted} should provide
further information on the binary nature of \os.

Both \hk\ (SD) and \hz\ (CC) were easily detected in the LORRI \onebyone images,
but neither appears to be a resolved binary based on our analysis.
Our observations rule out equal brightness binaries at separations $>$430~km for \hk\
and $>$260~km for \hz.
The CC KBO \pn\ was only marginally detected \mbox{(SNR $\approx$ 2.3-3.5)} in the
LORRI images, precluding definitive statements about its binarity.

In summary, at least one of four, and probably two of four, of the CC KBOs discussed here
are binaries.
Furthermore, these binaries have the tightest orbits of any resolved KBO binary systems,
demonstrating another unique capability of the \nh program in its exploration
of the Kuiper belt.
Our results extend to smaller bodies and to smaller semimajor axes the findings from previous studies 
\citep{fraser:2017, fraser:2017correction, noll:tno2020} that a significant
fraction ($\geq$20\%) of the CC KBO population is comprised of binaries.

\begin{acknowledgments}
This work was supported by NASA's  \nh project through contracts NASW-02008 and NAS5-97271/TaskOrder30.
We thank Susan Bennechi, Will Grundy, Alan Stern, and Anne Verbiscer for their detailed comments
on earlier drafts of the paper, and we thank Tod Lauer for independently confirming the non-PSF nature of
\jy\ and \os.
We thank the \nh Mission Operations team (especially Alice Bowman, Omar Custodio,
Helen Hart, and Karl Whittenburg), the \nh Science Operations team (Debi Rose, Nicole Martin, Ann Harch,
and Emma Birath), and the GNC lead (Gabe Rogers) for their expert planning and 
execution of these complicated observations.
We also thank two anonymous reviewers, whose suggested changes improved the paper. 
\end{acknowledgments}

%



\software{Interactive Data Language (IDL), licensed by the Harris Corporation,
SciPy \citep{virtanen:2020}, emcee package \citep{foreman:mcmc-2013}, 
HOTPANTS \citep{becker:2015}, spiceypy package \citep{annex:2020}.
}
\clearpage

\bibliography{bibtex}
\bibliographystyle{aasjournal}




\end{document}